# Reclaiming Constitutional Authority of Algorithmic Power

Yiyang Mei[1] & Michael J Broyde[2]

Abstract

Whether and how to govern AI is no longer a question of technical regulation. It is a question of constitutional authority. Across jurisdictions, algorithmic systems now perform functions once reserved to public institutions: allocating welfare, determining legal status, mediating access to housing, employment, and healthcare. These are not merely administrative operations. They are acts of rule. Yet the dominant models of AI governance fail to confront this reality. The European approach centers on rights-based oversight, presenting its regulatory framework as a principled defense of human dignity. The American model relies on decentralized experimentation, treating fragmentation as a proxy for democratic legitimacy. Both, in different ways, evade the structural question: who authorizes algorithmic power, through what institutions, and on what terms. This Article offers an alternative. Drawing from early modern Reformed political thought, it reconstructs a constitutional framework grounded in covenantal authority and the right of lawful resistance. It argues that algorithmic governance must rest on three principles. First, that all public power must be lawfully delegated through participatory authorization. Second, that authority must be structured across representative communities with the standing to consent, contest, or refuse. Third, that individuals retain a constitutional right to resist systems that impose orthodoxy or erode the domain of conscience. These principles are then operationalized through doctrinal analysis of federalism, nondelegation, compelled speech, and structural accountability. On this view, the legitimacy of algorithmic governance turns not on procedural safeguards or policy design, but on whether it reflects a constitutional order in which power is authorized by the governed, constrained by law, and answerable to those it affects.

---

[1] Yiyang Mei is an SJD candidate at Emory University, where she focuses on the intersection of law and technology. She holds a JD and MPH in epidemiology, also from Emory.
[2] Michael J. Broyde is a Professor of Law at Emory University, the Berman Projects Director in its Center for the Study of Law and Religion.

**Table of Contents**

## INTRODUCTION

Across both sides of the Atlantic and in democracies far and wide, the question of how to govern AI has become a stand-in for deeper debates over legitimacy, authority, and the institutional ordering of technological power.[3] Two schools of thought now dominate. The first, centered on the European Union, casts rights-centered regulation as the gold standard. As illustrated in Professor Anu Bradford's theory of the Brussels Effect, this view holds that as firms across the globe adopt instruments such as the General Data Protection Regulation and the AI Act, they are not merely complying with legal standards but affirming Europe's normative vision—a vision grounded in rights, dignity, and human oversight.[4] This reading, though influential, demands reconsideration.[5] It mistakes market-driven adaptation for moral assent. What appears as voluntary alignment is more often compelled imitation with regulation, driven by fear of regulatory exclusion. Worse still, it treats European legal values—liberty, privacy, autonomy—as neutral design principles, fit for export regardless of political tradition or institutional context. In doing so, it repeats a familiar civilizational script: a universalism that cloaks European particularity in the language of human rights.

The second school grounds itself in American federalism.[6] According to this account, the current patchwork of state-level AI initiatives—ranging from California's audit mandates[7] to Illinois's biometric privacy statutes[8]—reflects a deeper commitment to decentralization, local experimentation, and pluralism.[9] What may appear as fragmentation is reframed as innovation. In this telling, the absence of a comprehensive federal statute is not a failure but a feature.[10] It signals that democratic legitimacy is being approximated through institutional diversity. But here too, the account obscures more than it reveals. What is described as pluralism often masks a deeper void. Algorithmic systems are routinely deployed without legislative authorization, public deliberation, or civic consent. Decisions once made by human officials are now outsourced to automated systems designed and implemented by private actors. Individuals have no genuine means of refusal. Their rights are triggered only after the harm occurs. And the institutions that claim to speak on their behalf often lack the jurisdictional standing to intervene.

---

[3] Dean Woodley Ball, *How to Regulate Artificial Intelligence*, NATIONAL AFFAIRS (Spring 2024), https://nationalaffairs.com/publications/detail/how-to-regulate-artificial-intelligence. *See e.g.,* Cary Coglianese & Lavi M. Ben Dor, *AI in Adjudication and Administration*, 86 BROOK. L. REV. 791, 792 (2021) ("This article seeks to capture the state of the art in current uses of digitization, algorithmic tools, and machine learning in domestic governance in the United States."); Aziz Z. Huq, *A Right to a Human Decision*, 160 VA. L. REV. 611, 651 (2020) [hereinafter Huq, A Right to a Human Decision] ("I focus on the direct state applications of machine-learning tools to individuals for the purpose of allocating benefits or burdens."); Ryan Calo & Danielle Keats Citron, *The Automated Administrative State: A Crisis of Legitimacy*, 70 EMORY L.J. 797, 800 (2021) (noting the trend in state and federal public benefits agencies towards incorporating automated systems).
[4] *See infra* Part I. A.
[5] *Id.*
[6] *See infra* Part I. B.
[7] *See* Assemb. B. 1405, 2025-2026 Leg., Reg. Sess. (Cal. 2025), https://legiscan.com/CA/text/AB1405/id/3201524.
[8] *See* 740 ILCS 14/10.
[9] *Supra* note 4.
[10] *Id.*

Both models, in different ways, mischaracterize the nature of the systems they seek to regulate. Technically, algorithmic systems are tools.[11] In practice, they perform core public functions: allocating benefits,[12] processing immigration claims,[13] assessing criminal risk,[14] and regulating access to housing.[15] These are not incidental tasks. They are acts of governance and ruling. And yet, the systems that perform them operate with little oversight, minimal statutory grounding, and no clear line of constitutional accountability.[16] The first step in governing AI, then, is to name it for what it is: an instrument of public power. From there, the harder questions follow. Who governs through these systems? Under what authority? And with what structural safeguards?

This Article offers an answer by turning away from the dominant paradigm of liberal proceduralism and toward a constitutional tradition largely forgotten in contemporary debates: the early modern Reformed tradition of covenantal authority, consociation-al structure, and lawful resistance.[17] This tradition does not begin with abstract rights or administrative rationality. It begins with mutual obligation and collective authorship. It insists that those subject to power must authorize its exercise. That authority must be shared across representative communities, distributed and coordinated among plural bodies that retain their own institutional identity and jurisdictional claims.

The framework that emerges is grounded in three principles. Covenant affirms that algorithmic governance, like all public authority, must rest with the people and their representative institutions.

---

[11] e.g., they augment human decision making. *See* Chuck Brooks, *Augmenting Human Capabilities with Artificial Intelligence Agents*, FORBES, (July 31, 2024), https://www.forbes.com/sites/chuckbrooks/2024/07/31/augmenting-human-capabilities-with-artificial-intelligence-agents/; David De Cremer and Garry Kasparov, *AI Should Augment Human Intelligence, not Replace It*, HARVARD BUSINESS REVIEW (March 18, 2021), https://hbr.org/2021/03/ai-should-augment-human-intelligence-not-replace-it.

[12] *See* Michael Lokshin and Nithin Umapathi, *AI for Social Protection: Mind the People*, BROOKINGS (February 23, 2022), https://www.brookings.edu/articles/ai-for-social-protection-mind-the-people/.

[13] *See* Steven Hubbard, Invisible Gatekeepers: DHS' Growing Use of AI in Immigration Decisions, AMERICAN IMMIGRATION COUNCIL (May 9, 2025), https://www.americanimmigrationcouncil.org/blog/invisible-gatekeepers-dhs-growing-use-of-ai-in-immigration-decisions/.

[14] *See DOJ Report on AI in Criminal Justice: Key Takeaways*, COUNCIL ON CRIMINAL JUSTICE (April 2025), https://counciloncj.org/doj-report-on-ai-in-criminal-justice-key-takeaways/.

[15] For an overview of AI's use in housing, *see AI and Housing*, ALGORITHMIC JUSTICE LEAGUE, https://www.ajl.org/harms/housing (last visited July 15, 2025).

[16] For an overview of human oversight issue in AI, *see* Sarah Sterz et al., On the Quest for Effectiveness in Human Oversight: Interdisciplinary Perspectives (unpublished manuscript) (onfile with ArXiv)

[17] Consociation is a means of achieving self-determination for communities in pluralist or deeply divided places. It's especially appropriate where the residential mixing of populations make outright sovereign independence for each community inn their own nation-state- or territorial autonomy for each community in their own regional state. Originally, it was developed by jurist and philosopher Johannes Althusius, who sought to co-operation and co-existence among the cities of post-reformation Germany. Then, in the 20th century, the Dutch political scientist Arend Lijphart revived the term to describe political systems in which parallel communities, differentiated by ethnicity, language, or culture, share political power while retaining autonomy in matters of profound concern to them. At its core, it recognizes that political unity need not require cultural uniformity. It rejects the liberal fiction of the undivided citizenry and instead build institutions that allow multiple groups to govern together without surrendering their distinct identities. This is done through structured pluralism: power divided among groups, decisions made through negotiated agreement, and each community retains meaningful control over issues vital to its moral and cultural life - education, language, religion, family, and internal governance. The goal is not to eliminate difference but to give it political form. For a more complete explain, *see* Brendan O'Leary, *Consociation*, ENCYCLOPEDIA PRINCETONIENSIS, https://pesd.princeton.edu/node/246 (last viisted Aug 6, 2025)

It holds that AI systems must be lawfully delegated, morally bounded, and accountable to those they govern. That states may not deploy systems that intrude upon the domain of conscience. That federal agencies may not compel or preempt local authority. And that any system performing public functions must remain subject to constitutional review. Consociation affirms that governance must be structured across multiple planes: Horizontally, it recognizes the standing of tribal governments, tenant councils, and other local institutions to authorize or reject the use of algorithmic tools within their own domains; vertically, it limits federal coercion by treating funding conditions as constitutional contracts, which local jurisdictions may lawfully decline. Resistance marks the outer limit of lawful power. It protects the right of individuals to refuse submission to systems that encode dominant moral assumptions under the guise of neutrality. That protection does not depend on actual harm. A credible threat of interference is enough.

The purpose of this inquiry is not to build a new theory of AI exceptionalism, but to recover the constitutional logic that already governs the delegation of public authority in American law and apply it to AI. Algorithmic systems, when used by the state, must be held to the same standards that apply to any instrument of public rule. They must be authorized through law, structured through representative institutions, and constrained by a constitutional framework that protects local autonomy, institutional pluralism, and the right of refusal.

The Article proceeds in three parts. Part I critiques the prevailing models of AI governance for failing to recognize algorithmic systems as infrastructures of governance. It argues that both the European rights-based model and the American decentralization approach evade the structural questions of legitimate authority, community authorship, and constitutional accountability. Part II proposes an alternative grounded in the principles of covenant, consociation, and resistance. Drawing from Reformation political thought and American constitutional doctrine, it reconstructs the terms on which algorithmic governance may be exercised lawfully. Part III concludes by acknowledging the limits of this inquiry. This is not a comprehensive settlement of every doctrinal issue, nor could it be. It is, instead, a first-order intervention that seeks to shift the conversation—from audits and ethics to authority and law. From regulation as technical management to governance as a constitutional problem. From the language of rights to the structure of rule.

## Part I. The Failure of Current AI Governance Frameworks

This part criticizes the two main schools of AI governance. The first is the *Brussels Effect* and its associated literature, most notably Professor Anu Bradford's claim that EU law has become a global standard through normative admiration.[18] We argue that what this body of literature takes as voluntary alignment is often compliance without consent: risk management masked as moral leadership, with the result being a subtle reenactment of Euro-centric worldview, that the European regulatory framework can be exported wholesale under the banner of human rights, regardless of the political and legal traditions of other regimes. The second is the American federalism approach, one in which the state-level initiatives are often praised as laboratories of democratic experimentation. As this Section shows, both schools fail to address the structural question of legitimate authorization, such as what kind of authority is being exercised, through what traditions, and on behalf of whom? By reposing these questions, this Part lays the groundwork for a constitutional reconstruction of AI governance grounded in principles of covenant, consociation, and lawful resistance.

---

[18] *See generally* ANU BRADFORD, BRUSSELS EFFECT (Oxford U. Press, March 2020).

## A. Compliance Without Legitimacy: The Myth of the Brussels Effect

The first school of AI governance is what Professor Anu Bradford has termed the *Brussels Effect*: a theory that began as a descriptive account of regulatory diffusion and has since hardened into a normative claim about rightful rule.[19] First articulated in her book The Brussels Effect, Professor Bradford contends that the EU, by virtue of its vast market size and regulatory stringency, exports its legal standards beyond its borders through a kind of economic gravitation.[20] Corporations, faced with the high costs of compliance and the risk of exclusion from the EU's lucrative market, frequently adopt European rules preemptively, even in jurisdictions where those rules carry no formal legal force.[21] This, for Bradford, becomes evidence that European values merit global dominance: that when firms or governments conform, they are not simply managing risk, but affirming the normative superiority of Europe's rights-based regulatory model.[22] And by implication, those jurisdictions that do not reflect these values, or whose legal orders depart from the EU's liberal consensus, are cast as backward.[23] Since then, this narrative has largely set the tone for subsequent scholarship.[24] Yet it leaves several conceptual gaps, as this section argues: First, it confuses compliance with legitimate authority; second, it assumes the universality of European liberal values; third, it adopts a Euro-centric view that has long been criticized in international law.

### 1. Mistaking Compliance as Legitimate Authority.

---

[19] For instance, the following literature has adopted similar stance. *See* Ryan Calo, *Artificial Intelligence Policy: A Primer and Roadmap.* 51 U.C. DAVIS L. REV. 399 (2017) (discussing the U.S. market-oriented approach to AI regulation, where innovation and economic growth are prioritized, often at the expense of individual rights and protections seen in other jurisdictions); *See also* Mark MacCarthy, *Fairness in Algorithmic Decision-Making*, BROOKINGS (December 6, 2019), https://www.brookings.edu/articles/fairness-in-algorithmic-decision-making/ (exploring differences in fairness approaches, noting that U.S. AI regulation often leans toward market-driven solutions and self-regulation); Joshua A. Kroll, Joanna Huey, Solon Barocas, Edward W. Felten, Joel R. Reidenberg, David G. Robinson & Harlan Yu, *Accountable Algorithms*, 165 U. PA. L. REV. 633 (2017) (discussing that U.S. data privacy regulations are characterized by a fragmented, sectoral approach with decentralized regulatory authority. It indicates that the US approach, focused on sector-specific laws and enforcement by agencies like the FTC, doesn't provide a comprehensive privacy framework, and often prioritizes market interests and commercial flexibility over broad data protection rights).
[20] *Supra* Note 18.
[21] *Id.*
[22] *Id.*
[23] Of course, the text itself didn't explicitly use the term "backward," but the contrast drawn is unflattering in the sense that it suggests that while China may have technological prowess, it doesn't set global standards because its regulatory regime lacks credibility and normative authority, especially in areas like data protection and content moderation.
[24] *Supra* Note 19.

The first problem with the argument in *Brussels Effect* as described above is that it treats voluntary compliance as rightful authority.[25] Take Meta, which reengineered its global data infrastructure to comply with the EU's General Data Protection Regulation (GDPR) by introducing consent mechanisms, minimizing data retention, and localizing European user data;[26] or consider Google, which restructured its advertising model and search algorithms to satisfy EU's antitrust rulings and the Digital Markets Act by unbundling services, revising ranking logic, and offering browser choice screen[27] – both corporations change their practices not out of the executives' admiration for liberal principles, but that noncompliance is costly. Meta, for instance, derives roughly a quarter of its global revenue from Europe, and has already been fined over €2.5 billion under the GDPR, including a record €1.2 billion penalty in May 2023 for unlawful data transfers.[28] Google, too, has adjusted its systems to avoid penalties that can reach 4% of global turnover, sanctions that now routinely exceed €20 million per infraction.[29] The calculus is simple: obey or pay.

But Professor Bradford seems to read this differently. She treats the international corporations' global uptake of European frameworks as confirmation of their moral superiority, that when firms conform to EU rules, they are not merely managing legal and financial risk. They are, in her words, "validating [the EU's] regulatory agenda."[30] What begins as the corporations' strategic adaptation becomes, in her telling, normative assent. The underlying logic appears to be: if a rule is widely followed, and if it carries the formal trappings of rights, such as proportionality, explainability, or contestability, then such adoption is presumed just; others' compliance, regardless of their legal traditions or political cultures, is taken as evidence of a kind of legitimacy that is automatically conferred.[31]

Such a presumption is most explicit in Bradford's noteworthy Article *Europe's Digital Constitution*.[32] In that essay, she argues that instruments such as the GDPR, the Digital Services Act, and the AI Act together "engrains Europe's human-centric, rights-preserving, democracy-enhancing, and redistributive vision into binding law."[33] Once they are adopted by foreign firms or national regulators, their legitimacy is taken as self-evident, because the rules incorporate procedural safeguards and rights-like features.[34] **But that is not how legitimacy works.** Legitimacy requires more than rule-following.

---

[25] *Supra* Note 18.
[26] For a report of how Meta has changed in compliance with EU laws, *see* Luis Rijo, *Meta Confronts EU Watchdog on "Beyond the Law" Compliance Demands*, PPC LAND (Mar 8, 2025), https://ppc.land/meta-confronts-eu-watchdog-on-beyond-the-law-compliance-demands/.
[27] *See* Naomi Buchanan, *Google Announces Changes for EU Users to Comply with the DMA*, INVESTOPEDIA (March 05, 2024), https://www.investopedia.com/google-announces-changes-for-eu-users-to-comply-with-the-dma-8604230.
[28] *See 1.2 Billion Euro Fine for Facebook as a Result of EDPB*, EUROPEAN DATA PROTECTION BOARD (May 22, 2023), https://www.edpb.europa.eu/news/news/2023/12-billion-euro-fine-facebook-result-edpb-binding-decision_en.
[29] *See* Kristof Van Quathem, Anna Sophia Obserschelp de Meneses, and Nicholas Shepherd, *Google Fined €50 Million in France for GDPR Violation*, COVINGTON (Jan 22, 2019), https://www.insideprivacy.com/eu-data-protection/google-fined-e50-million-in-france-for-gdpr-violation/.
[30] *Supra* Note 18 at 89.
[31] *Id.*
[32] *See generally* Anu Bradford, *Europe's Digital Constitution*, 64 VA. J. INT'L L. 1 (2023)
[33] *Id.*
[34] *Id.*

It demands consent by the governed.[35] As Max Weber puts it in *Economy and Society*, rational-legal authority does not arise from rules alone.[36] It depends on "a belief in the legality of enacted rules and in the right of those elevated to authority under such rules to issue commands."[37] In other words, the GDPR's authority, if it is to be legitimate, cannot rest on the fact that it is followed. It must rest on recognition that the EU, as its issuer, holds rightful power to govern those who comply, including international firms such as Meta and Google, and foreign jurisdictions such as Brazil or Kenya. Yet this is precisely what is not happening. When these actors adopt EU rules, their compliance is less rooted in allegiance to European legal traditions, they have their own; and more plausibly driven by risk aversion: an effort to avoid regulatory penalties or exclusion from the European market. In this light, the diffusion of EU rules is less normative assent by the governed than jurisdictional overreach.

Other philosophers have likewise confirmed this point. Lon Fuller, for instance, in *The Morality of Law*, argued that legality is not a matter of form alone, but of fidelity to a moral enterprise.[38] Law, he insisted, is not merely a structure of command, but "the enterprise of subjecting human conduct to the governance of rules" – an enterprise that demands more than formal intelligibility or administrative efficiency, but requiring that the governed recognize the laws' ends as legitimate.[39] Similarly, Joseph Raz, in his "service conception" of authority, advances this claim, that an authority is legitimate only when it helps its subjects better conform to reasons they already have.[40] In this sense, legitimacy is not grounded in constraint or convenience, but in normative service.[41] Robert Cover, in *Nomos and Narrative*, pushes this point further still: "No set of legal institutions or prescriptions exists apart from the narratives that locate it and give it meaning."[42] Law, for Cover, cannot be separated from the communities and traditions that interpret it, or else it ceases to govern and becomes mere command.[43]

In this sense, the conclusion is clear: foreign corporations' compliance with European regulations does not reflect recognition of the EU's normative authority, but rather a rational response to the threat of market exclusion.

### 2. Assumption of the universality of European values

The second mistake of this argument is to assume the universality and exportability of European liberal values. It goes like this: because Europeans advocate rights and freedoms, which are undeniably indispensable to democratic life, and many states have voluntarily adopted similar frameworks, these values must be universal.[44] After all, who would object to a domestic framework that protects freedom and human dignity?

---

[35] *See* Leslie Green, *Law, Legitimacy, and Consent*, 62 S. CAL.L. REV. 795, 795 (1989).
[36] *See* Stephen Kalberg, *Max Weber's Types of Rationality: Cornerstones for the Analysis of Rationalization Processes in History,* 85 AM.J.SOC. 1145,1161 (1980)
[37] *See* MAX WEBER, ECONOMY AND SOCIETY Vol. 1, at 215 (Guenther Roth & Claus Wittich eds., Ephraim Fischoff et al. trans., Univ. of Cal. Press 1978) (1922).
[38] *See generally* Lon L. Fuller, *Positivism and Fidelity to Law: A Reply to Professor Hart*, 71 HARV. L. REV 630 (1958)
[39] *See* LON FULLER, MORALITY OF LAW 2 (Yale University Press, 1969).
[40] *See generally* Joseph Raz, *The Problem of Authority: Revisiting the Service Conception* 90 MINN L. REV 1003 (2006).
[41] *Id.*
[42] *See* Robert Cover, *Foreword: Nomos and Narrative,* 97 HARV. L.R 1, 4 (1983)
[43] *Id.*

But concepts like liberty, privacy, and autonomy are not universal.[45] They are historical constructs forged in particular contexts, in response to context-sensitive political crises, particularly in their secondary forms.[46] To treat them as abstract, exportable goods would be to commit a form of normative anachronism and collapses intellectual history into ideology, replacing the contingencies of conflict with the illusion of consensus. Put plainly, when states adopt European values in the name of freedom and democracy, irrespective of their own legal, political, and historical traditions, they are less affirming the universality of European values than to absorb thoughts born out of foreign crises, detached from the cultural and institutional contexts that gave them meaning. Indeed, compliance often masquerades as consensus. When, under pressure from President Trump, NATO member states pledged to increase their defense spending to meet the 2 percent GDP threshold, the move appeared to reflect agreement with a nationalist vision of burden-sharing.[47] In truth, it was but a concession to political and economic leverage driven by imperative to preserve alliance cohesion and to avoid retaliation. To read such alignment as evidence of normative convergence is to mistake diplomacy for conviction and to confuse the choreography of power with the architecture of belief.

Take *libertas*, for instance. In contemporary AI governance literature, it is often repackaged as a universal right. Digital ethicists and legal scholars, including Bradford herself, treat liberty and autonomy as normatively self-evident and globally desirable.[48] In *The Brussels Effect*, Bradford lauds EU law for its strong commitment to rights-based values, celebrating its defense of individual freedom and dignity in the digital age.[49]

But *libertas* is not a universal abstraction. It arose from particular crises in Western Europe and carries a distinct genealogy.[50] In the mid-14th century, Bartolus of Sassoferrato defined *libertas* as a form of privilege – the right of city-states like Florence and Siena to preserve their own laws and customs within a nominal empire.[51] In the 15th century, with the fall of the Florentine republic and the return of Medici rule, Machiavelli reimagined *libertas* in *Discourses on Livy* as the capacity for collective self-rule through institutional contestation.[52] Then, in 17th-century England, due to its constitutional crisis, this republican ideal of liberty reemerged. James Harrington, in *The Commonwealth of Oceana*, and Algernon Sidney, in his *Discourses*, recast *libertas* as the right of a people to frame their own laws through representative assemblies.[53] Based on this intellectual history, it is obvious that each political rupture gave the term a new meaning.[54] *Libertas* has never been a stable, self-evident good gradually unfolding over time.[55] It has always been contingent, shaped by conflicts and forged through power struggles. To treat it as a universal value fit for export is to sever it from its context, and in doing so, vacates the historical memory embedded in such a concept.[56]

---

[44] *Supra* Note 18.
[45] *See* Daniel J. Solove, *Conceptualizing Privacy* 90 Cal.L.Rev. 1087, 1093 (2002)
[46] *See generally* QUENTIN SKINNER, *Liberty Before Liberalism* (Cambridge Univ. Press 1998).
[47] *See* David Vergun, *NATO Leaders Pledge to Increase Defense Spending*, U.S. DEPARTMENT OF DEFENSE (June 25, 2025), https://www.defense.gov/News/News-Stories/Article/Article/4226009/nato-leaders-pledge-to-increase-defense-spending/.
[48] *See* Carina Prunkl, *Human Autonomy at Risk? An Analysis of the Challenges from AI*, 34 MINDS & MACHINES 26 (2024), https://doi.org/10.1007/s11023-024-09665-1; *see also* Neil Renic and Elke Schwarz, *Inhuman-in-the-Loop: AI-targetting and the Erosion of Moral Restraint*, OPINIOJURIS( Dec. 19, 2023), http://opiniojuris.org/2023/12/19/inhuman-in-the-loop-ai-targeting-and-the-erosion-of-moral-restraint/.
[49] *Supra* Note 16 at 157 (case studies of major European legislations that uphold rights and human dignity)
[50] For a report on the idea of liberty and why it has flourished in the West, *see* Jim Powell, *Why has Liberty Flourished in the West?*, CATO POLICY REPORT (Sep 2020), https://www.cato.org/sites/cato.org/files/serials/files/policy-report/2000/9/liberty-flourished.pdf.
[51] QUNETIN SKINNER, THE FOUNDATION OF MODERN POLITICAL THOUGHT VOL.1 at 12, (Cambridge Univ. Press 1978) (elaborating how he uses the concept to vindicate in legal terms the claims made by the cities about their liberty).

This same logic applies to the discourse of rights. In contemporary AI governance literature, proposals for new rights are multiplying at warp-speed.[57] Scholars now slap a new right onto every harm supposedly caused by AI: because AI is a black box, there must be a right to explanation;[58] because it profiles individuals, there must be a right to opt out.[59] But these proposals miss the historical fact that rights are, largely, context-bound claims articulated within concrete constitutional struggles, and against identifiable modes of domination.

For instance, during the Protestant Reformation, reformers like John Ponet, in *A Short Treatise of Political Power* (1556), argued that rulers who violated God's law or the natural order forfeited their authority and could be lawfully resisted.[60] Such a right was formulated as a response to Mary I's violent suppression of Protestantism, and the broader danger of rulers invoking divine right to justify arbitrary power.[61] A generation later, the Huguenot tract *Vindiciae contra tyrannos* (1579) declared that subjects not only had the right, but the duty, to oppose monarchs who breached the sacred covenant between God, ruler, and people.[62] The rights to depose a tyrant and to preserve one's conscience against state orthodoxy were responses to religious violence, political persecution, and existential contests over sovereignty.[63] To speak of new rights detached from such political foundations and to treat them instead as abstract remedies to moral discomfort would be risking reducing them to empty gestures that are largely meaningless.

---

[52] *See generally* QUNETIN SKINNER, LIBERTY BEFORE LIBERALISM (Cambridge Univ. Press 1998)
[53] *Id.*
[54] *See generally* QUNETIN SKINNER, VISIONS OF POLITICS VOL.1 REGARDING METHOD 57-91 (Cambridge Univ. Press 2002) (explaining there is no such thing as "dateless wisdom;" ideas are to be found in the contexts that gave them meaning).
[55] *Id*. at 178 (summarizing that he has written in his book *Liberty Before Liberalism* the rise and fall within Anglophone political theory of a particular view about the concept of liberty).
[56] For a critique of the academic practice of taking concepts as dateless wisdom, *see supra* Note 66.
[57] For example, there have been proposals for Rights to a Human Decision Maker, Right to Disconnect, Right to a Human-to-Human Interaction, etc., among others. *See From Digital Rights to International Human Rights: The Emerging Right to a Human Decision Maker*, INSTITUTE FOR ETHICS IN AI (Dec. 11, 2023), https://www.oxford-aiethics.ox.ac.uk/blog/digital-rights-international-human-rights-emerging-right-human-decision-maker; Carl De Cicco, Aselle Ibraimova, Alison Heaton & Claudia Gwinn, *AI in the Workplace – is Regulation on its Way in the UK?*, REEDSMITH (June 10, 2024), https://www.employmentlawwatch.com/2024/06/articles/employment-uk/ai-in-the-workplace-is-regulation-on-its-way-in-the-uk/; *A Right to a Human-to-Human Interaction*, INSTITUTE FOR ETHICS IN AI (Nov 7, 2024), https://www.oxford-aiethics.ox.ac.uk/blog/right-human-human-interaction.
[58] *See e.g*., Lilian Edwards & Michael Veale, *Enslaving the Algorithm: From a "Right to an Explanation" to a "Right to Better Decisions"?*, 16 IEEE SECURITY & PRIVACY 46 (2018); Bryce Goodman & Seth Flaxman, *European Union Regulations on Algorithmic Decision-Making and "a Right to Explanation"*, 38 AI MAG. 50, 55–56 (2017); Gianclaudio Malgieri & Giovanni Comandé, *Why a Right to Legibility of Automated Decision-Making Exists in the General Data Protection Regulation*, 7 INT'L DATA PRIVACY L. 243, 246 (2017); Antoni Roig, *Safeguards for the Right Not to Be Subject to a Decision Based Solely on Automated Processing (Article 22 GDPR)*, 8 EURO. J. L. & TECH. 1 (2017); Sandra Wachter, Brent Mittelstadt & Chris Russell, *Counterfactual Explanations without Opening the Black Box: Automated Decisions and the GDPR*, 31 HARV. J. L. & TECH. 841 (2018).
[59] *See* Antonio Cordella and Francesco Gualdi, *Regulating Generative AI: The Limits of Technology-Neutral Regulatory Frameworks. Insights from Italy's Intervention on ChatGPT*, 41 GOV'T INFO. Q. 101982 (2024); Marta Williams et al., *Ethical Data Acquisition for LLMs and AI Algorithms in Healthcare*, 7 NPJ DIGIT. MED. No.377 (2024).
[60] *See* John Ponet, *Short Treatise on Political Power*, CATO INSTITUTE, https://cdn.cato.org/libertarianismdotorg/books/ShortTreatiseonPoliticalPower.pdf (last visited Jul 15, 2025).

### 3. The Discredited Eurocentric perspective

The third mistake of this argument is its Eurocentric view of AI regulation that has been largely discredited in international law for its violent pedigree. Historically, the notion that a polity must emulate European norms in order to be admitted to the international family of nations has a long history.[64] In 1530, for instance, Francisco de Vitoria justified Spain's conquest of the Americas under the guise of natural law.[65] Cloaked in the language of civility and salvation, Vitoria argued that indigenous peoples who refused Christian preaching or resisted free trade could be lawfully compelled to submit: "Barbarians are obliged to accept the faith of Christ… if they are asked to do so and refuse, in the law of war action may be taken against them."[66] In his thought, the indigenous tribes' refusal to conform to European religious, commercial, and political order nullified their native sovereignty.[67] They must adopt the European order for their society to be recognized as legitimate.[68]

Then, in the 18th and 19th century, European jurists repackaged legal domination in the language of Enlightenment. Emer de Vattel, in *The Law of Nations* (1758), asserted that nations that are still savage cannot be recognized as full members of the international community unless they embraced commerce, cultivation, and European forms of governance.[69] John Stuart Mill, in *A Few Words on Non-*

---

[61] *See* David M. Whitford, *John Adams, John Ponet, and a Lutheran Influence on the American Revolution*, LUTHERAN QUAR. Vol. XV 143, 146 (2001).
[62] *See generally* Kathleen W. MacArthur, *The Vindiciae Contra Tyrannos: A Chapter in the Struggle for Religious Freedom in France*, 9 CHURCH HIS. NO.4 285 (1940).
[63] *Id.*
[64] *See* Andrew Linklater, *The "Standard of Civilization" in World Politics*, 5 Social Character, Historical Processes (2016), https://quod.lib.umich.edu/h/humfig/11217607.0005.205?view=text;rgn=main (explaining that the European powers insisted that only "civilised" states could belong to international society and be subject to international law… the family of nations displayed some of the hallmarks of a society "consisting of persons interested in maintaining the rules of good breeding" and committed to "shunning intercourse with those who do not observe them.")
[65] *See* Antony Anghie, *Francisco de Vitoria and the Colonial Origins of International Law*, 5 SOC. & LEGAL STUD. 326, 326 (1996).
[66] *See* ON THE AMERICAN INDIANS 231, 265, https://warwick.ac.uk/fac/arts/history/students/modules/archive/hi3f9/timetable/spanishinventionofrightsandinternationallaw/on-the-american-indians.pdf?utm_source=chatgpt.com (last visited Jul 16, 2025).
[67] *Id.*
[68] *Id.*
[69] *See* EMER DE VATTEL, THE LAWS OF NATIONS Book 1 Ch.7 §81 (stating that the cultivation of the soil is an obligation imposed by nature … those nations who inhabit fertile countries but disdain to cultivate their lands … are injurious to all their neighbors, and deserve to be extirpated as savage and pernicious beasts… and therefore their conquest is extremely lawful).

*Intervention* (1859), drew a stark line between "civilized" and "barbarous" peoples,[70] justifying despotism in the latter case, as they "have no rights as a nation."[71] James Lorimer, a founding member of the Institut de Droit International, went further. In 1883, he proposed a formal hierarchy of legal recognition based on civilizational capacity, insisting that international law must exclude races of lesser value.[72]

To see how this logic works in practice, consider the overthrow of the Hawaiian monarchy as an example. In 1893, a coalition of American sugar planters and business elites organized under the Committee of Safety staged a coup against Queen Lili'uokalani with the support of U.S. Marines dispatched from the USS Boston.[73] This newly formed provisional government, composed almost entirely of non-native residents, installed a political regime under the 1887 Bayonet Constitution.[74] In this document, the planters replaced the kingdom's constitutional framework with Anglo-American legal institutions, imposed property and income qualifications for voting, and codified commercial laws designed to facilitate foreign investment and land acquisition.[75] While these reforms were, officially, neutral instruments of legal modernization; in effect, they dismantled indigenous sovereignty and entrenched a settler regime that redefined lawful governance in terms of racial exclusion, commercial utility, in alignment with Euro-American norms.

Judged against this background, it becomes very difficult to deny that the same structure might be reappearing in the global diffusion of AI regulations. Once again, under the banners of "human rights," "human-centric design," and "digital dignity," a new form of extraterritorial legal order is quietly taking hold. Brazil's Bill No. 2338/2023, for instance, approved by the Senate in December 2024, share many vocabularies with the EU AI Act.[76] Both frame AI governance around the dignity and centrality of the human person;[77] both place rights-based constraints on AI deployment;[78] both emphasize that AI must not determine democratic institutions or the legality of governance;[79] both share a concern over algorithmic bias, especially toward vulnerable groups.[80] Yet, it is not clear how much of "Brazil" is left: there's no mentioning of collective or national sovereignty over data infrastructure; AI infrastructure is also not taken as sites of jurisdictional or territorial control. It's not clear if alignment with EU values is truly promoting fairness and human rights, or simply that countries with less global market power needs to emulate EU in order to secure access to market at the cost of native sovereignty.

---

[70] *See* John Stuart Mill, *A Few Words on Non-Intervention*, 27 New Eng. Rev 252, 259 ("to suppose that the same international customs, and the same rules of international morality, can obtain between one civilized nation and another, and between civilized nations and barbarians, is a grave error.")
[71] *Id.*
[72] *See* Martti Koskenniemi, *Race, Hierarchy and International Law: Lorimer's Legal Science*, 2 EJIL 27 415, 421 (e.g., in *the Institutes of Law,* he enlisted general jurisprudence against the studies of primitive culture that were spreading with developmental anthropology at the time, "for our purposes, the single life of Socrates is of greater value than the whole existence of the Negro race.")
[73] *See Jan.17, 1893 | Hawaiian Monarchy Overthrown by America-Backed Businessmen*, THE NEW YORK TIMES (Jan. 17,1893), https://archive.nytimes.com/learning.blogs.nytimes.com/2012/01/17/jan-17-1893-hawaiian-monarchy-overthrown-by-america-backed-businessmen/.
[74] *Id.* Named such due to the duress used to ratify it.
[75] DAVIANNA POMAIKA'I MCGREGOR & MELODY KAPILIALOHA MACKENZIE, OFFICE OF HAWAIIAN AFFAIRS, MO'OLELO EA O NA'HAWAI'I: HISTORY OF NATIVE HAWAIIAN GOVERANCE IN HAWAI'I at 297; *see also The 1887 Bayonet Constitution: The Beginning of the Insurgency*, HAWAIIAN KINGDOM BLOG (Aug 25, 2014), https://hawaiiankingdom.org/blog/the-1887-bayonet-constitution-the-beginning-of-the-insurgency/.
[76] *See* Lais Martins, *Brazil's AI Law Faces Uncertain Future as Big Tech Warms to Trump*, TECH POLICY PRESS (Feb 4, 2025), https://www.techpolicy.press/brazils-ai-law-faces-uncertain-future-as-big-tech-warms-to-trump/#.
[77] *See e.g.,* Regulation (EU) 2024/1689 of the European Parliament and of the Council of 13 June 2024, 2024 O.J. (L 2024) 1 (6) (laying down harmonised rules on artificial intelligence) (hereinafter "EU AI Act"); Projeto de Lei No. ___, de 2023, Dispõe sobre o uso da Inteligência Artificial (Braz.). Art. 2.I (hereinafter "Brazil AI Act").
[78] EU AI Act (1); Brazil AI Act Art. 40.

## B. US Federalism without foundational questions

The second school of AI governance refers to the literature on American federalism that notes the patchwork of state regulations in the absence of a federal statute.[81] According to this view, the states' ongoing enactment of AI rules in line with their distinct priorities reflects a structural feature of the American tradition.[82] It is said to embody a spirit of innovation, experimentation, and pluralism.[83] States, acting without instruction from the federal government, begin a generative process of regulation that brings in labor unions, civil society, technical experts, and public stakeholders to debate the meaning of AI safety.[84]

Yet, this literature has not addressed several foundational questions. Who authorized the delegation of governance functions to automated systems? When these systems determine eligibility for benefits or access to public services without meaningful human and institutional oversight, are they engaged in state action in the constitutional sense? If they are, do they implicate procedural protections under the Due Process Clause? Are First Amendment freedoms being chilled when private systems deny services based on unclear or ideologically charged moderation rules? More fundamentally, what kind of authority is being exercised, under what legal theory, and with what safeguards for liberty? It is to these questions that the next Part turns.

---

[79] EU AI Act (1); Brazil AI Act Art. 1.
[80] EU AI Act (7); Brazil AI Act Chap.2, Art.7.
[81] *See* James C. Cooper and Evangelos Razis, *The Federalist's Dilemma: State AI Regulation & Pathways Forward*, Harv. J.L. & Pub. Pol'y (forthcoming) (noting that while states acting as laboratories is a good thing, when the subject of regulation is interstate, this patchwork of state regime is far from ideal); David Rubenstein, *Federalism & Algorithms* 67 Ariz. L. Rev 1 (forthcoming); Kristin O'Donoghue, *A Patchwork of State AI Regulation is Bad. A Moratorium is Worse*, AI FRONTIERS (June 25, 2025), https://ai-frontiers.org/articles/congress-might-block-states-from-regulating-ai; Kristian Stout, *State Approaches to AI Regulation are a Patchwork*, INTERNATIONAL CENTER FOR LAW & ECONOMICS (June 10, 2025), https://laweconcenter.org/resources/state-approaches-to-ai-regulation-are-a-patchwork/.
[82] *Id.* For instance, as Rubenstein notes, federal law provides little protection against political deepfakes. The Federal Election Campaign Act of 1971 primarily addresses campaign finance regulations rather than content manipulation. The Act contains no provisions explicitly addressing synthetic media, and its broader provisions on campaign communications do not clearly encompass AI-generated content intended to deceive voters. Meanwhile, at least 20 states enacted political deepfakes in the runup to the 2024 elections. Their methods are different: while some impose criminal sanctions, other impose civil liability; some confer agency officials with exclusive enforcement authority, whereas others allow private rights of action.
[83] *Id.*
[84] Scott Kohler, *Technology Federalism: U.S. States at the Vanguard of AI Governance*, CARNEGIE ENDOWMENT (February 10, 2025), https://carnegieendowment.org/research/2025/02/technology-federalism-us-states-at-the-vanguard-of-ai-governance?lang=en.

## PART II. NEW ALGORITHMIC GOVERNANCE FRAMEWORK

This Part proposes a new framework for algorithmic governance, grounded in the Reformation tradition that later shaped the American constitutional order.[85] It has four Sections. Section A explains why it is necessary to return to these traditions. Section B examines how the Constitution, consistent with the covenantal order, requires that the authority to regulate AI must remain with the people. Section C articulates a consociational order in which algorithmic governance is authorized horizontally and vertically. Section D concludes that individuals have a structural right to resist when they are subject to credible threat of punishment.

### A. Why Returning to This Tradition

It is important to return to these traditions because the principles of covenant, consociation, and resistance each form an essential aspect of the constitutional framework without which such an order would lack justification, structure, and constraint.

**1. Covenant offers five structural promises of governance that current regulations fail to secure: participatory authorization, moral grounding, lawful delegation, equitable distribution, and the recognition of moral personhood**

---

[85] The principles are drawn from the Reformation tradition because it articulated one of the earliest constitutional frameworks in which political authority is derived from covenant. Through Puritan federal theology and the political theory of thinkers such as Johannes Althusius, Theodore Beza, and John Calvin, this tradition shaped Anglo-American constitutionalism. E.g., Calvinists emphasized liberty of conscience ad free exercise of religion; the rights were extended to include the rights to assemble, worship, educate, contract, associate ad publish; Calvinist communities Geneva introduced the separation of church and state while coordinating them through shared moral ends; there also emerged congregational elections for church officers, local synods and transparent procedural rules as early prototypes of constitutional government emphasizing the rule of law, popular representation, and procedural fairness. Puritan New England became the Constitutional seedbed. In Massachusetts and elsewhere, for instance, Calvinist ideas took legal forms in documents like the Body of Liberties and Massachusetts Constitution. These texts enumerated religious, procedural, and economic rights, and institutional checks on political and ecclesiastical power. Additionally, the notion that legitimate authority arises from the people's covenant, and that rulers are trustees and accountable to that compact, became central. John Milton extended this to argue that every person is a prophet, priest, and king, with rights to speak, worship and govern. This laid intellectual groundwork for democratic participation and civil liberties. *See* JOHN WITTE JR., THE REFORMATION OF RIGHTS: LAW, RELIGION, AND HUMAN RIGHTS IN EARLY CALVINISM (Cambridge Univ. Press 2007); cf. ERIC NELSON, THE HEBREW REPUBLIC: JEWISH SOURCES AND THE TRANSFORMATION OF EUROPEA POLITICAL THOUGHT (Harv. Univ. Press 2010)

Covenant offers five structural promises of governance: (participatory) authorization by the governed,[86] moral grounding in shared norms,[87] lawful delegation by the people,[88] equitable standing of the subjects,[89] and the recognition of moral personhood.[90] As the article will show, these promises correspond to four constitutional principles: First, AI must be governed by the people—and by the representative communities authorized to speak on their behalf.[91] Second, domains of conscience lie beyond the reach of the state.[92] Third, federal agencies may not compel states to adopt particular AI systems, nor may they preempt states' sovereign authority to regulate such systems within their own jurisdictions.[93] Fourth, AI governance must ensure the equal standing of all subjects.[94] The remainder of this section focuses on elaborating the five covenantal promises.

Begin with participatory authorization. Although the Reformation tradition didn't use this vocabulary, it articulated a political theory in which authority must be authorized by those subject to such power. As Theodore Beza wrote, magistrates are put into their offices by the election or consent of the people.[95] More systematically, the *Vindiciae Contra Tyrannos*, an influential Huguenot tract published in Basel that undergirded the Dutch Revolt against Spanish rule, argues that political authority rests on two interlocking covenants: one divine, between God, king, and people; and one human, between king and people alone.[96] The dual convent structure secures that power is not unilateral, but publicly conferred, grounded in the people's original authority to condition rule.

Of course, authorization by the public alone is not enough. A just polity must also involve a shared commitment to the common good.[97] As Althusius writes, "the symbiotes are co-workers who, by the bond of an associating and uniting agreement, communicate among themselves whatever is appropriate for a comfortable life of soul and body. In other words, they are participants in a common life."[98] Cicero likewise echoed this view, calling the political community "a gathering of men associated by a consensus as to the right and a sharing of what is useful."[99]

---

[86] *See e.g.,* John Witte Jr., *Rights, Resistance, and Revolution in the Western Tradition: Early Protestant Foundations*, 26 LAW AND HISTORY REVIEW 67, 78 (explaining that for Beza, the people's right to vote for or consent to their rulers was essential to the legitimacy of the political regime… magistrates are put into their offices by the election or the consent of the people); *see also* Junius Brutus, Vindiciae Contra Tyrannos at 16, https://www.yorku.ca/comninel/courses/3020pdf/vindiciae.pdf (last visited July 17, 2025) (contending that political authority depends on two intertwined pacts: one divine such as God-king-people, and one social, that between king and people. Sovereignty holds only so long as both parties keep their promises.)
[87] *See* JOHANNES ALTHUSIUS, POLITICA, Chap.I (as he says, politics is the art of associating men for conserving social life where members pledge mutual sharing of goods, labor and justice by explicit or tacit agreement)
[88] *Id* at chapter VI-VII.
[89] *Id* at XIX.
[90] *See* JOHN CALVIN, THE INSTITUTES OF THE CHRISTIAN RELIGION Book 1, Chapter 15 (saying that the conscience is not the property of pope, prince, or parliament, but belongs to God).
[91] *Infra* Part II. B.1
[92] *Id*. B.2
[93] *Id*. B.3
[94] *Id.* B.4
[95] *Id.*
[96] *Id.*
[97] *See generally* Maximilian Jaede, *The Concept of the Common Good* (University of Edinburgh Working Paper) 1 – 18 https://www.thebritishacademy.ac.uk/documents/1851/Jaede.pdf

The limits of this polity life are residents' conscience.[100] Polities, grounded in shared obligation of residents, may not violate what belongs to the person by nature.[101] As Althusius cautions, religion is not to be commanded by force but to be instructed by the Word.[102] One ought to retain "the liberty to know, to utter, and to argue freely according to conscience,"[103] as "the care of each man's soul belongs unto himself."[104] All such ideas laid essential groundwork for what would later emerge in the American constitutional tradition as the First Amendment's protection of conscience.[105]

Public authority is upheld through lawful delegation by the people.[106] By consent, citizens entrust specific offices with the administration of shared goods and duties.[107] Public office thereby becomes a site of fiduciary responsibility, exercising power within the bounds of the purposes and terms for which it was granted.[108] As Althusius puts it, "the things done by the senatorial collegium are considered done by the whole community that the collegium represents."[109] In this sense, authority flows from below and remains accountable to those who conferred it.

Within this structure, all individuals stand as equals, recognized as moral persons by nature, equally capable of participating in public life, protected by law, and bound by reciprocal rights and duties in pursuit of the common good.[110] To note, this is not a call for egalitarian redistribution, as but a recognition that natural and vocational diversity is part of the providential ordering of society.[111]

---

[98] *Supra* Note 83.
[99] *Id.*
[100] Particularly in Calvinist and Huguenot traditions, civil authority could not intrude upon the internal forums of conscience.
[101] *Supra* Note 84.
[102] *See* POLITICA Ch.XXVIII
[103] *See* John Milton, Aeropagitica: For the Liberty of Unlicenc'd Printing, to the Parliament of England
[104] *See* John Locke, A Letter Concerning Toleration, Amendment I (Religion)
[105] *See* Michael J. White, *The First Amendment's Religion Clauses: "Freedom of Conscience" Versus Institutional Accommodation*, 47 SAN DIEGO L.REV 1075, 1075 (2010) (arguing that though the phrase, "freedom of conscience" is not to be found in the Constitution, the then-contemporary Protestant conception of freedom of conscience was presupposed in these two clauses."); *see also* John Witte Jr., *Back to the Sources? What's Clear and Not So Clear About the Original Intent of the First Amendment*, 47 BYU L.REV 1303,1308 (2022); Steven D. Smith, *What Does Religion Have to do with Freedom of Conscience*, 76 U.COLO. L. REV 911 (2005).
[106] JOHN LOCKE, SECOND TREATISE OF GOVERNMENT, Ch.11, §141 ("the legislature cannot transfer the power of making laws to any other hands. It was delegated to them from the people, and they aren't free to pass it on to others. Only the people can decide the form of the commonwealth, which they do by instituting a legislature and deciding whose hands to put it into…")
[107] POLITICA Ch. XIX (explaining what is the Supreme Magistrate, that is he who, having been constituted according to the laws of the universal association for its welfare and utility, administers its rights and commands compliance with them).
[108] *Id.*
[109] *Id* at Ch. VI.

Together, these aspects form the structural foundation for governance: that power be authorized from the ground up through common consent, such authority be exercised among equals, each person's conscience remains beyond the reach of rule, and that the polity be ordered toward the pursuit of shared goods.

### 2. Consociation offers a constitutional alternative to AI federalism by grounding the authority to govern in representative communities

Consociation offers an alternative to AI federalism by grounding the authority to govern in representative communities. Current accounts of AI federalism place heavy emphasis on state experimentation, as if pluralism and accountability might emerge from the spontaneous diffusion of subnational initiatives.[112] While this framework has practical appeal, it fails to address the deeper structural problem: how to give individuals, and by extension their representative communities, the standing to contest algorithmic decisions that shape their lives but do not reflect their voice?

Consociation offers a second path. Chronologically, this structure is older than federalism.[113] It goes beyond the federal-state governmental relationship and instead, names a symbiotic association – a covenantal union of distinct communities which, without coercion or command, bind themselves into a common polity entrusted with the duty to protect the needs and norms of the members they represent. In the context of AI, this may exactly be what is needed.[114]

---

[110] As opposed to data points and passive subjects waiting to be rescued. Much of contemporary AI governance literature frames individuals as passive targets of systematic risks, unable to change their environments according to their preferences. Yet, people keep shows agency by curating digital footprints, scrubbing their profiles in anticipation of search results, and learning to game the logic of the algorithm. The question seems less, as the mainstream literature asks, if people are harmed? Or, if their privacy is violated, but rather, are they given spaces to reason and dissent?

[111] In contemporary and secular terms, it means that what we are pursuing is equality of opportunity, rather than equality of outcomes, as disparities in result, by themselves, do not necessarily mean injustice or exclusion. As Thomas Sowell notes, racial disparities sometimes reflect deeper structural conditions such as family formation, educational background, geographic location or cultural orientation, rather than overt or systemic discrimination.

[112] *Supra* Note 78-81.

[113] Federalism as a modern constitutional arrangement emerges, most explicitly, in the 18th century, with the US Constitution as the paradigmatic example. "Federal," etymologically, derives from *foedus*, Latin for "covenant," "treaty", "compact," or "alliance." Its institutional division of powers between a national and subnational government with dual sovereignty is a distinctly modern invention, responding to Enlightenment theories of sovereignty and post-Westphalian state. Consociation as a political form, however, traces back to the Reformed tradition, particularly in the works of thinkers like Althusius in the early 17th century, as mentioned above. *See* W. West Allen, *President's Message: The Constitution Empowers Us*, FEDERAL BAR ASSOCIATION (January 29, 2021), https://www.fedbar.org/blog/presidents-message-the-constitution-empowers-us/; *see also* FEDERAL, ETYMONLINE, https://www.etymonline.com/word/federal (last visited Jul 18, 2025).

[114] *See generally* ETHICS OF ARTIFICIAL INTELLIGENCE, UNESCO, https://www.unesco.org/en/artificial-intelligence/recommendation-ethics (last visited Jul 18, 2025); *see also* Barbara DeLollis and Steven Melendez, *AI Expert: Transparency and Accountability Crucial to AI Governance*, HARVARD BUSINESS SCHOOL (May 29, 2025), https://www.hbs.edu/bigs/marc-rotenberg-artificial-intelligence.

### 3. Resistance establishes structural protections against credible threat of punishment

Resistance refers to the capacity of individuals to legally refuse submission to unjust or unlawful authority.[115] It is not rebellion in the anarchic sense, but a duty of fidelity to the covenant, a commitment to uphold the polity's founding terms when those in power violate them.[116] As *Vindiciae contra Tyrannos* puts it, when the king violates the covenant, and has not kept promise, the subject is no longer tied to them; the subject may lawfully resist.[117] How many roadblocks may be lawfully placed in the way of expanding illicit authority.

Of course, constitutional law today does not formally recognize a right to resist in the Reformation sense.[118] It does not treat disobedience as an act of fidelity, nor does it grant standing to contest the legitimacy of power beyond procedural defect.[119] Yet something of the structure remains: it still permits individuals to challenge state action based on a credible threat of punishment, thereby functioning as a form of ex-ante constraint.[120]

Indeed, this point is largely missed in contemporary AI governance literature. So far, the body of literature remains preoccupied with proximate concerns such as the erosion of privacy and individual agency.[121] Their arguments go as follows: because AI, by profiling people's online and offline behaviors, is capable of persuading and manipulating at scale, individuals must be protected from such technologies in order to retain autonomy,[122] primarily through notice, disclosure, and the asserting their right to explanation.[123]

---

[115] *See* VINDICIAE CONTRA TYRANNOS, Question 2.
[116] *Id.*
[117] *Id.*
[118] *See* Robert Goldwin, *Is There an American Right of Revolution?*, AMERICAN ENTERPRISE INSTITUTE (January 01, 1990), https://www.aei.org/articles/is-there-an-american-right-of-revolution/ (saying that while the Declaration of Independence is not a manifesto for revolution, it asserts a continuing political principle: the right of the people to alter or abolish non-governmental tyranny, not the right of individuals to revolt for its own sake); *see also generally* Tom Ginsburg, Daniel Lansberg-Rodriguez, and Mila Versteeg, *When to Overthrow your Government: The Right to Resist in the World's Constitutions* 60 UCLA L. REV. 1184 (2013).
[119] *Id.*
[120] *See infra*. Part 2. D.
[121] *See* Cornelia C. Walther, *Are You At Risk of Acute Agency Decay Amid AI?*, FORBES (March 29, 2025), https://www.forbes.com/sites/corneliawalther/2025/03/29/are-you-at-risk-of-acute-agency-decay-amid-ai/; Simone Di Plinio, *Panta Rh-AI: Assessing Multifaceted AI Threats on Human Agency and Identity* 11 SOC SCI. & HUMANS. OPEN 101434 (2025); Katherine Miller, *Privacy in an AI Era: How do we Protect Our Personal Information*, HAI (march 18, 2024), https://hai.stanford.edu/news/privacy-ai-era-how-do-we-protect-our-personal-information; David Elliot and Eldon Soifer, *AI Technologies, Privacy, and Security*, FRONT ARTIF INTELL. (2022); Kelly D. Martin and Johnna Zimmermanni, *Artificial Intelligence and Its Implication for Data Privacy*, 58 CURRENT OPINION PSYCHOL 101829 (2024).
[122] *See e.g.,* S.C. Matz, *The Potential of Generative AI for Personalized Persuasion at Scale*, 14 SCI. REP 4692 (2024); Lisa P. Argyle, Testing Theories of Political Persuasion Using AI 122 PNAS (2025); Nicola Davis, *AI Can be More Persuasive than Humans in Debates, Scientists Find*, THE GUARDIAN (May 19, 2025), https://www.theguardian.com/technology/2025/may/19/ai-can-be-more-persuasive-than-humans-in-debates-scientists-find-implications-for-elections.
[123] *See* Lkubisa Metikos, *The Right to an Explanation in Practice: Insights from Case law for the GDPR and the AI Act*, 17 LAW INNOV. & TECH 205 (2025); Margot E. Kaminski, Gianclaudio Malgieri, *The Right to Explanation in the*

But it is questionable whether privacy and agency, in the deep constitutional sense, are actually being lost. People do not lose their agency simply because they are influenced.[124] Historically, individuals have always been shaped by their environments.[125] Consider the church: for centuries in medieval Europe, it not only guided people's moral and spiritual choices, but ordered the very structure of life prescribing what to believe,[126] how to pray,[127] whom to marry,[128] which vocations were honorable,[129] and even how one ought to die.[130] Living under the influence of an external environment is but an inherited condition of our political life. It should not be correct to think that, just because this source of influence is switched, individual agency is at peril. What should instead be asked is, assuming such influence exists, whether it flows from the authorities they recognize, and if they possess meaningful ways to resist.

## B. Four Covenant Principles in Constitutional Law

This Section retrieves constitutional doctrines to elaborate four maxims that constrain the legitimate use of algorithmic systems by the state. First, the power to regulate AI must rest with the people, and by extension, with the representative communities that speak on their behalf. Second, states may not adopt or fund AI systems that intrude upon conscience or compel speech or belief. Third, federal agencies may not coerce states into adopting particular AI systems, nor preempt their authority to

---

*AI Act* 1-30 (U. COLO. L. LEGAL STUD. RES. PAPER NO.25-9) (2025), https://papers.ssrn.com/sol3/papers.cfm?abstract_id=5194301; Jarek Gryz, marcin Rojszczak, *Black Box Algorithms and the Rights of Individuals: No Easy Solution to the "Explanability" Problem* 10 INTERNET POL'Y REV. 2 (2021); Simona Demkova, *The AI Act's Right to Explanation: A Plea for an Integrated Remedy*, MEDIA LAWS(October 31, 2024), https://www.medialaws.eu/the-ai-acts-right-to-explanation-a-plea-for-an-integrated-remedy/.

[124] *See generally* Mustafa Emirbayer and Ann Mische, *What is Agency?* 103 AM.J. SOCIOL 962-1023 (1998) (arguing that agency shouldn't be treated as a vague, catch-all term or reduced to any one of its dimensions such as habit and intention, but should be conceptualized as a temporally embedded process – a structured engagement with time, shaped by the past, oriented toward the future, and responsive to the present. In this sense, influence is a condition of agency, not its negation); Janeen D. Loehr, *The Sense of Agency in Joint Action: An Integrative Review*, 29 PSYCHON. BULL. & REV 1089-1117 (2022) (summarizing that people have a sense of shared agency when they perform joint actions in pairs. When people collaborate with AI then, they develop a shared sense of agency with AI, rather than having theirs eroded).

[125] *See* Tyler Biscontini, *Person-in-Environment (PIE) Theory*, EBSCO, https://www.ebsco.com/research-starters/social-sciences-and-humanities/person-environment-pie-theory (last visited July 19, 2025); *see generally* Roger S. Gamble et al., *The Role of External Factors in Affect-Sharing and Their Neural Bases*, 157 NEUROSCI. & BIOBEHAV. REV 105540 (February 2024); Cephas Tetteh et al., *How do Environmental Factors Shape Entrepreneurial Intention? A Review and Future Research* 20 NATURE 2955-2977 (2024) (for investigating how external environments contribute to entrepreneurial intention); Jennifer Leeman et al., *Applying Theory to Explain the Influence of Factors External to an Organization on the Implementation of an Evidence-Based Intervention*, 2 FRONT. HEALTH SERV. 889786 (2022).

[126] The Church established orthodoxy through ecumenical councils such as the Council of Nicaea in 325.

[127] The Church structured daily prayer through the Liturgy of the Hours.

[128] The Church asserted jurisdiction over marriage law, including banns, consent, degrees of consanguinity, and sacramental validity. After the Decretum Gratiani and Fourth Lateran Council, marriage was sacramental and required public ritual in a church.

[129] E.g., the Church established the tripartite division of society, such as those who pray (oratores), those who fight (bellatores), and those who work (laboratories).

[130] The Church oversaw the *ars moriendi* (the art of dying) tradition, providing manuals, rituals and confessional preparation for death.

regulate such systems within their own jurisdictions. Fourth, when used by the state, AI systems must not discriminate, silence, exclude, or evade public accountability.

### 1. The power to regulate AI must ultimately rest with the people, and, by extension, with the representative communities that speak on their behalf.

The power to regulate AI must ultimately rest with the people and, by extension, with the representative communities authorized to speak on their behalf. As the Preamble declares, "we the people … do ordain and establish this Constitution."[131] Since it is the people that ordain the very framework through which public power is authorized, they should also govern AI as it exercises a form of public power.

That it is the people, not the states, and certainly not powerful corporations, that must wield the authority to govern AI is affirmed in constitutional doctrine. In *McCulloch v. Maryland*,[132] the Court addressed whether the state of Maryland could impose a tax on the federally chartered Bank of the United States.[133] Maryland contended that, because the Constitution was silent on the creation of banks, Congress lacked the authority to incorporate one.[134] But even if such a power existed, the state insisted that it retained sovereign authority to tax institutions operating within its borders.[135] As the counsel for the state of Maryland has argued, "the powers of the general government…are delegated by the states, who alone are truly sovereign; and must be exercised in subordination to the states, who alone possess supreme domination."[136]

The Court rejected that view. It held that the federal Constitution is not a compact among the states but a government deriving its legitimacy directly from the people.[137] "The government of the Union… is, emphatically and truly, a government of the people. In form and in substance it emanates from them. Its powers are granted by them, and are to be exercised directly on them, and for their benefit."[138] That power cannot be made dependent on the discretion of the states. "It has been said, that the people had already surrendered all their powers to the state sovereignties... But surely, the question whether they may resume and modify the power granted to the government does not remain to be settled in this country."[139]

To allow the states to tax the Bank, Marshall warned, would be to allow them to destroy it. "The power to tax involves the power to destroy."[140] And if states could destroy a federal instrument, they could destroy the very means by which the national government governs.[141] "If the states may tax one instrument, they may tax another; and thus they may tax all the means employed by the government... This was not intended by the American people. They did not design to make their government dependent on the states."[142] Although the case is, of course, not about AI, it does emphasize a foundational point that public power must remain accountable to its source, the people, and not be made contingent on the discretion of subordinate jurisdictions or private entities.

---

[131] *See* U.S CONST. PMBL.
[132] 17 U.S. 316 (1819).
[133] *Id* at 319.
[134] *Id* at 320-321.
[135] *Id.*
[136] *Id* at 403.
[137] *Id* at 401.
[138] *Id* at 404-405.
[139] *Id.*
[140] *Id* at 431.

How, then, do the people wield such power? Through representative communities authorized to act on their behalf. In the domain of labor, this includes unions, which serve as intermediaries between workers and managerial authority. In *Fibreboard Paper Products Corp. v. NLRB*,[143] for instance, Fibreboard unilaterally subcontracted maintenance work at its Emeryville, California plant, terminating unionized employees without prior negotiation.[144] The Court held that such a measure involves terms and conditions of employment, which falls at the very heart of the employment relationship, and is peculiarly suitable for resolution within the collective bargaining framework.[145]

The terms and conditions of employment subject to collective bargaining are not to be reshaped by administrative fiat. In *Colgate-Palmolive-Peet Co. v. NLRB*,[146] the employer discharged 37 workers under a closed-shop agreement after the union expelled them for supporting a rival organization.[147] The Board deemed the discharges unlawful, arguing that enforcing the agreement interfered with workers' Section 7 rights.[148] The Court rejected this view. It held that where the closed-shop contract was valid under both state law and Section 8(3) of the Act, and entered into in good faith, the employer could lawfully enforce its terms.[149] Administrative policy could not override the statutory protection afforded to a collectively bargained agreement.[150]

Of course, these cases are not about AI per se. But they clarify a principle: that the conditions shaping the terms of employment are subject to collective bargaining, and the results of that bargaining are not easily displaced. Since AI now structures the workplace by tracking productivity, assigning shifts, and flagging underperformance in real time,[151] its deployment properly falls within the union's

---

[141] *Id.*
[142] *Id* at 431-432.
[143] 379 U.S. 203 (1964)
[144] *Id* at 205.
[145] *Id* at 215 (however, the court also emphasized that this decision was limited to the specific facts of the case and didn't encompass all forms of subcontracting.)
[146] 338 U.S. 355 (1949)
[147] *Id* at 359.
[148] *Id.*
[149] *Id.* at 361
[150] *Id.* at 362-363
[151] *See* Amanda Downie and Molly Hayes, *AI in the Workplace: Digital Labor and the Future of Work*, IBM, https://www.ibm.com/think/topics/ai-in-the-workplace (last visited July 20, 2025); Hannah Mayer, Lareina Yee, Michael Chui and Roger Roberts, *Superagency in the Workplace: Empowering People to Unlock AI's Full Potential*, MCKINSEY DIGITAL, https://www.mckinsey.com/capabilities/mckinsey-digital/our-insights/superagency-in-the-workplace-empowering-people-to-unlock-ais-full-potential-at-work (last visited July 20, 2025); Luona Lin and Kim Parker, *U.S. Workers are More Worried Than Hopeful About Future AI Use in the Workplace*, PEW RESEARCH CENTER

domain.¹⁵² The outcomes of such bargaining must be respected. They should not be overridden by administrative agencies issuing unilateral implementation guidelines for AI systems. In this sense, when unions negotiate the terms under which AI is deployed, they act as representatives of the people; and through them, the people govern the systems that govern their work. Other domains likely follow suit.¹⁵³

**2. The states' and other public institutions' adoption, funding, and deployment of AI systems must be categorically limited where such use intrudes upon the domain of conscience.**

The adoption, funding, or deployment of AI systems by states and public institutions must be categorically restricted where such use encroaches upon the domain of conscience. This domain includes not only the inward sphere of individual conviction, but also the communal structures through which conscience is formed: familial, religious, educational, and cultural ways of life. These are pre-political authorities, rooted in traditions that both precede and constrain the reach of public power. Their integrity calls for recognition and must be met with respect, not subjected to algorithmic assimilation. To note, the cases cited here are framed primarily in terms of religious liberty, but the point is less to assert a particular religious exemption, but to recognize that certain spheres of life - those through which conscience is cultivated and authority made intelligible - are not open to technological substitution: there are limits to what may be delegated, and conscience draws that line.¹⁵⁴

---

(February 25, 2025), https://www.pewresearch.org/social-trends/2025/02/25/u-s-workers-are-more-worried-than-hopeful-about-future-ai-use-in-the-workplace/.

¹⁵² *See generally* Bradford Kelley, *Belaboring the Algorithm: Artificial Intelligence and Labor Unions*, YALE J. ON REGUL. BULL (June 15, 2024); *see also* Aurelia Glass, *Unions Give Workers a Voice Over How AI Affects Their Jobs*, CAP (May 16, 2024), https://www.americanprogress.org/article/unions-give-workers-a-voice-over-how-ai-affects-their-jobs/; Caroline Burnett, JT Charron, Nadege Dallais, Elizabeth Ebersole, Rachel Farr, Matthias Kohler & Autumn Sharp, *Navigating Labor's Response to AI: Proactive Strategies for Multinational Employers Across the Atlantic*, THE EMPLOYER REPORT (June 20, 2025), https://www.theemployerreport.com/2025/06/navigating-labors-response-to-ai-proactive-strategies-for-multinational-employers-across-the-atlantic/.

¹⁵³ For example, in public housing, city councils might represent residents to govern. *See* Jashayla Pettigrew and Lisa Balick, *Hurting Everybody: Portland City Council, Tenant Sound off on Proposed AI Rent-Fixing Software Ban*, KOIN (April 2, 2025), https://www.koin.com/local/hurting-everybody-portland-city-council-tenants-sound-off-on-proposed-ai-rent-fixing-software-ban/; Emily Davis, *Jersey City Just Banned Landlords from Using AI to Set Rent – a First for the Garden State*, NEW YORK POST (May 23, 2025), https://nypost.com/2025/05/23/real-estate/jersey-city-council-just-banned-the-use-of-ai-to-set-rent/.

¹⁵⁴ For families as pre-political communities, *see* Meyer v. Nebraska 262 U.S. 390 (1923) (recognizing that the family has a fundamental right to direct the upbringing and education of children, implying that the family isn't merely a creature of the state but has prior standing); Pierce v. Society of the Sisters of the Holy Names of Jesus and Mary 268 U.S. 510 (1925) (affirming that the family has a fundamental right to direct the upbringing and education of children; noting that the child is not the mere creature of the state). For religious groups' authority beyond the reach of public power, *see* Hosanna-Tabor Evangelical Lutheran Church and School v. E.E.O.C 565 U.S. 171 (2012) (court recognizing a "ministerial exception" grounded in the autonomy of religious institutions) *See also e.g.,* Government Relations, UNITED STATES CONFERENCE OF CATHOLIC BISHOPS, https://www.usccb.org/offices/government-relations#:~:text=Breadcrumb,policy%20departments%20at%20the%20USCCB (last visited Aug 6, 2025). and much more on this problem.

That individuals may not be coerced into affirming beliefs contrary to their convictions is affirmed in *West Virginia State Board of Education v. Barnette*.[155] In that case, public school students in Minersville, West Virginia were required by state regulation to salute the American flag and recite the Pledge of Allegiance, under penalty of expulsion.[156] Jehovah's Witness families objected, arguing that the ritual constituted idolatry and violated their religious convictions.[157] The Court struck down the requirement, holding that compelled patriotic expression violated the First Amendment's guarantees of free speech and free exercise of religion.[158] As Justice Jackson wrote: "If there is any fixed star in our constitutional constellation, it is that no official, high or petty, can prescribe what shall be orthodox in politics, nationalism, religion, or other matters of opinion."[159] This is the first case in which Supreme Court squarely recognized that such freedom of conscience includes not only the right to speak but also the right not to be made a mouthpiece for state orthodoxy.[160]

In addition to individual conscience, communities also have the right to shape their moral worlds apart from the assimilating designs of the state. In *Wisconsin v. Yoder*, Amish families challenged a compulsory schooling law that required children to remain in school until the age of sixteen.[161] They argued that extended exposure to secular education threatened the integrity of their religious community and violated their obligation to withdraw from the modern world.[162] The Court sided with the families, holding that the Free Exercise Clause protected their right to direct their children's education in accordance with their faith.[163] "The values of parental direction of the religious upbringing and education of their children," the Court affirmed, "have a high place in our society."[164] Families are not instruments of public planning.[165] They are pre-political communities of moral formation—sites of conscience that both precede and limit the reach of state authority.[166]

Even when the state must engage with moral or religious claims, it must do so with restraint. In *Masterpiece Cakeshop v. Colorado Civil Rights Commission*,[167] a Christian baker, Jack Phillips, declined to create a custom wedding cake for a same-sex couple, citing his religious belief that marriage is a sacred union between a man and a woman.[168] He argued that being compelled to create the cake would violate his First Amendment rights to free speech and free exercise of religion.[169] The Colorado Civil Rights Commission charged him with unlawful discrimination under the state's public accommodations law.[170] But during the proceedings, several commissioners expressed open contempt for Phillips's religious views, with one stating that "freedom of religion and religion has been used to justify all kinds of discrimination throughout history, whether it be slavery, whether it be the Holocaust."[171] Phillips sued, arguing that the commission's bias violated his right to free exercise.[172] The Supreme Court agreed. It held that the Commission's treatment of Phillips violated the Free Exercise Clause.[173] Writing for the Court, Justice Kennedy emphasized that the Constitution commits government itself to religious tolerance, and that upon even slight suspicion that the state's hostility or bias has infected the proceedings, the guarantee has been violated.[174] Public authority, the Court made clear, should approach religious and moral claims with neutral and respectful consideration.[175] It may not ridicule or denigrate those beliefs, even in the name of equality.[176]

---

[155] 319 U.S. 624 (1943)
[156] *Id* at 626 – 628.
[157] *Id* at 629.
[158] *Id* at 642.
[159] *Id*.
[160] Before *Barnette*, the First Amendment was primarily interpreted as a shield against prior restraint or censorship. *See e.g.,* Near v. Minnesota , 283 U.S. 697 (1931)
[161] 405 U.S. 205, 207 (1972).
[162] *Id* at 211 and 212.
[163] *Id* at 235.
[164] *Id* at 214.
[165] *Id.*
[166] *Id.*
[167] 584 U.S. 617 (2018).
[168] *Id* at 626.
[169] *Id* at 630

Of course, these cases are not about AI. But they embody a principle that individuals have a right to be free from state interference in matters of belief and conscience, and that communities, whether religious congregations, tribal nations, and moral associations rooted in tradition, have a corresponding right to preserve their internal norms of life and meaning. Yet both forms of freedom are now being quietly eroded. Public institutions, including school districts and state agencies are funding and deploying AI systems for explicitly formative purposes such as providing moral guidance and determining access to public goods.[177] Several school districts, for example, have introduced AI mental-health chatbots, funded in part by public health grants, to promote emotional regulation, reduce behavioral disruptions, and improve academic performance among thousands of middle- and high-school students.[178]

---

[170] *Id* at 622.
[171] *Id* at 635.
[172] *Id* at 630.
[173] *Id* at 625.
[174] *Id* at 638.
[175] *Id.*
[176] *Id.*
[177] *See AI Technology is Helping Crisis Line Responders*, VA NEWS (September 2, 2024), https://news.va.gov/133911/ai-technology-is-helping-crisis-line-responders/; Sally Barksdale, *Veterans Mental Health Crisis: The Case for Artificial Intelligence Solutions*, HUHPR, http://www.huhpr.org/original-content/2025/4/6/veterans-mental-health-crisis-the-case-for-artificial-intelligence-solutions (last visited July 21, 2025); *See generally* Cynthia Geppert, *The Year of AI: Learning With Machines to Improve Veteran Health Care*, 41 FED PRACT. 388-389 (Dec 16, 2024); *The Government and Public Services AI Dossier*, DELOITTE, https://www.deloitte.com/us/en/what-we-do/capabilities/applied-artificial-intelligence/articles/ai-dossier-government-public-services.html (last visited July 21, 2025); Sanam Hooshidary, Chelsea Canada and William Clark, *Artificial Intelligence in Government: The Federal and State Landscape* (November 22, 2024), https://www.ncsl.org/technology-and-communication/artificial-intelligence-in-government-the-federal-and-state-landscape; *Artificial Intelligence in the States: Harnessing the Power of AI in the Public Sector,* THE COUNCIL OF STATE GOVERNMENTs (December 5, 2023), https://www.csg.org/2023/12/05/artificial-intelligence-in-the-public-sector-how-are-states-harnessing-the-power-of-ai/.
[178] *See* Julie Jargon, *When There's No School Counselor, There's a Bot*, THE WALL STREET JOURNAL (Feb. 22, 2025); *AI Therapy for Schools*, ABBY, https://abby.gg/schools/ (last visited July 21, 2025).

In this light, the question is not how to better align algorithmic systems with liberal values such as choice, transparency, or procedural fairness.[179] The deeper question is this: how does an algorithm that simulates moral formation differ, in kind, from the forms of compulsory service the Constitution already forbids the state from imposing? Where do we draw the line between voluntary exposure and state-sponsored coercion, especially when the system is designed to persuade rather than merely inform? And if such tools are capable not only of suggestion but of shaping belief, redirecting judgment, and subtly remaking moral identity, what follows for our understanding of freedom?

A full exposition of these questions, and their doctrinal consequences, lies beyond the scope of this paper. But to begin asking them is already a step toward recovering the proper boundaries of conscience and the constitutional order that protects it.

### 3. Federal agencies may not coerce states into using particular AI systems, nor preempting their authority to regulate them within their own jurisdictions.

Federal agencies may not compel states to adopt particular AI systems, nor may they preempt state authority to regulate such systems within their own jurisdictions, whether by federal mandate, conditional funding, or regulatory threat. Under the Constitution, that power is not theirs to claim. The Necessary and Proper Clause permits Congress to enact laws that carry into execution its enumerated powers,[180] but not to enlarge them.[181] And the Tenth Amendment draws the boundary in clear terms: "The powers not delegated to the United States by the Constitution, nor prohibited by it to the States, are reserved to the States respectively, or to the people."[182]

*Printz v. U.S.* made clear that Congress may not compel state or local executive officials to implement or enforce federal regulatory programs without their consent.[183] Jay Printz and Richard Mack, county sheriffs from Montana and Arizona, challenged provisions of the Brady Handgun Violence Prevention Act that required state and local law enforcement to conduct background checks on firearm purchasers until a federal system could be established.[184] The Court held that such a mandate violated the anti-commandeering principle embedded in the structure of the Constitution and confirmed by the Tenth Amendment.[185] The reason, the Court explained, is that even a law enacted pursuant to an enumerated power, such as the Commerce Clause, must still be proper under the Necessary and Proper Clause.[186] A law that conscripts state officers into federal service intrudes upon the sovereignty reserved to the states and therefore cannot be "proper" within the meaning of Article I.[187] As the Court put it, when a federal law violates the principle of state autonomy, it is not a proper means of carrying federal power into execution; it is, quoting The Federalist No. 33, "merely an act of usurpation."[188]

---

[179] Marie Christin Decker, Laila Wegner and Carmen Leicht-Scholten, *Procedural Fairness in Algorithmic Decision-Making: The Role of Public Engagement*, 27 ETHICS & INFO. TECH 1 (2025); Niamh Kinchin, *"Voiceless": the Procedural Gap in Algorithmic Justice*, 32 INT'L J.L. & INFO. TECH (2024); Michele Loi, Andrea Ferrario, Eleonora Vigano, *Transparency as Design Publicity: Explaining and Justifying Inscrutable Algorithms*, 23(3) ETHICS INF TECHNOL. 253-263 (2020)
[180] U.S. CONST. Art. I § 8, cl.18
[181] *See* United States v. Comstock 560 U.S. 126, 133 (1949) (saying that the clause grants Congress the authority to enact laws that are rationally related to the implementation of a constitutionally enumerated power).
[182] U.S. CONST. Amend X
[183] 521 U.S. 898, 935 (1997)
[184] *Id* at 904.
[185] *Id* at 933.
[186] *Id* at 923.

The federal government may also not coerce states into adopting federal programs through indirect financial compulsion. In *National Federation of Independent Business v. Sebelius*, Congress sought to expand Medicaid eligibility under the Affordable Care Act by conditioning continued Medicaid funding on state compliance: either accept the new terms or lose all existing Medicaid funds.[189] At the time, Medicaid accounted for over 10% of most state budgets.[190] The Court held that the federal government may not attach new conditions to existing funding in such a way that states have no meaningful choice but to comply.[191] This kind of financial compulsion, he wrote, was "a gun to the head."[192] "The Constitution simply does not give Congress the authority to compel the States to enact or enforce a federal regulatory program.[193] Even Congress's spending power—its most flexible and far-reaching tool—must operate within the bounds of lawful consent.[194] To threaten fiscal collapse as the price of noncompliance is coercion; lawful governance by consent does not secure obedience by threatening ruin.[195]

Furthermore, *Murphy v. NCAA* emphasized that Congress may not forbid states from refraining to act.[196] In that case, the State of New Jersey sought to repeal its own laws prohibiting sports betting.[197] But the Professional and Amateur Sports Protection Act (PASPA) barred any state from "sponsor[ing], operat[ing], advertis[ing], promot[ing], licens[ing], or authoriz[ing] by law" any form of sports gambling.[198] The Court struck down PASPA, holding that it violated the Constitution's federal structure by subordinating state sovereignty to federal command.[199] Writing for the majority, Justice Alito reasoned, Congress cannot issue direct orders to the governments of the States.[200] The federal government may regulate individuals under its enumerated powers.[201] But it may not conscript the states to carry out its will.[202]

---

[187] *Id.*
[188] *Id.*
[189] 567 U.S. 519, 539 -541 (2012).
[190] *Id*. at 542.
[191] *Id.*
[192] *Id.* at 581.
[193] *Id.* at 521.
[194] *Id.*
[195] *Id.*
[196] 584 U.S. 453, 474 (2018).
[197] *Id* at 464.
[198] *Id.*
[199] *Id.* at 486.
[200] *Id.* at 471.
[201] *Id.*
[202] *Id.*

Yet in promoting the development of AI, the federal government appears to be edging precisely toward commandeering the states. In 2024, the Senate committee on technology policy proposed allocating $500 million annually to states for AI and Automated Decision System (ADS) development, but only on the condition that states suspend all AI regulation through 2035.[203] Around the same time, Senator Ted Cruz introduced legislation that would withhold federal broadband funding from any state that attempts to regulate AI technologies within its own jurisdiction.[204]

Superficially, these proposals might appear to fall within the scope of ordinary fiscal incentives permitted under *South Dakota v. Dole*.[205] But broadband funding today underwrites essential state responsibilities, including expanding access in underserved regions, supporting telehealth, enabling public education, and fostering economic development.[206] Since 2021, Congress has appropriated over $65 billion for broadband expansion, including $42.45 billion through the BEAD program, structured as block grants to states for high-speed infrastructure.[207] To threaten the withdrawal of these funds unless states abandon AI regulation places them in a bind so severe that the choice is no real choice at all, but what the Court in *NFIB* called a "gun to the head."[208]

Other questions remain: At what point does federal guidance cease to be guidance and become de facto law? When federal funds are conditioned on silence or compliance, what remains of state sovereignty? And when regulatory authority is consolidated in executive agencies, without legislative enactment or local consent, are we not witnessing the quiet erosion of federalism itself? A full treatment of these questions lies beyond the scope of this section, but they are worth being posed now-- before the lines grow too blurred to recover.

---

[203] *See Senate Leaders Water Down 10-Year State AI Law Ban, Instead Tie it to Federal Funding Promise*, FISHER PHILLIPS (June 6, 2025), https://www.fisherphillips.com/en/news-insights/senate-leaders-water-down-10-year-state-ai-law-ban.html;

[204] *See Senate Upholds 10-Year State AI Law Ban in Trump's Budget Bill*, PERPLEXITY (June 22, 2025) https://www.perplexity.ai/page/senate-upholds-10-year-state-a-DtWwhrGTRIyHFpQo4Elg0Q; Loreben Tuquero, *"One Big Beautiful Bill" Could Block AI Regulations for 10 Years, Leaving its Harms Unchecked*, POLITIFACT (June 17, 2025), https://www.politifact.com/article/2025/jun/17/state-AI-laws-moratorium-reconciliation-bill/.

[205] 483 U.S. 203 (1987) (the Supreme Court held that Congress' use of its spending power to encourage states to adopt a minimum drinking age of 21 by conditioning the receipt of federal highway funds is a valid exercise of its spending power; Congress has the power to attach conditions to the receipt of federal funds to further broad policy objectives, as long as those conditions are in pursuits of the general welfare, unambiguous, related to a federal interest, and not barred by other constitutional provisions).

[206] *See Broadband Infrastructure Program*, NATIONAL TELECOMMUNICATIONS AND INFORMATION ADMINISTRATION, https://www.ntia.gov/funding-programs/high-speed-internet-programs/broadband-infrastructure-program (last visited July 22, 2025); *Addressing Broadband to Improve Access to Telehealth*, TELEHEALTH.HHS.GOV, https://telehealth.hhs.gov/community-stories/addressing-broadband-improve-access-telehealth (last visited July 22, 2025); Rachel Hirsch and Jake Varn, *Broadband Access for Success in Postsecondary Education*, NATIONAL GOVERNORS ASSOCIATION (April 7, 2021), https://www.nga.org/news/commentary/broadband-access-for-success-in-postsecondary-education/.

[207] Stephanie Weiner, Judson Cary, Legal and Compliance Support for the 42.45B Broadband Equity, Access, and Deployment (BEAD) Grant Program, NATIONAL ASSOCIATION OF ATTORNEYS GENERAL (June 25, 2024), https://www.naag.org/attorney-general-journal/legal-and-compliance-support-for-the-bead-grant-program/.

[208] 567 U.S. at 521.

1. **State Use of AI Must Meet Constitutional Standards of Equality and Due Process**

AI systems, when used by the state, must have their outcomes reviewed for discriminatory effects. Their decisions must afford individuals a meaningful opportunity to be heard. And their protections must extend to all persons, not merely citizens. States must be held responsible for statutory and constitutional violations caused by the systems they adopt, fund, or deploy.

Results from algorithmic systems must not materially discriminate across racial, religious, status, or community lines. In *Yick Wo v. Hopkins*, the Court struck down a facially neutral ordinance after more than 200 Chinese applicants were denied laundry permits while nearly every white applicant was approved.[209] The Equal Protection Clause, the Court made clear, forbids not only discriminatory laws but also discriminatory enforcement, especially when administrative discretion becomes a tool of exclusion.[210] "Though the law itself be fair on its face," the Court held, "if it is applied and administered by public authority with an evil eye and an unequal hand," it is unconstitutional.[211] By the same logic, a superficially neutral algorithmic system cannot be excused if it produces discriminatory results.[212] This is even more so problematic in the case of AI systems which are rarely understood enough to be clearly neutral – the black box problem. It is not enough to audit training data or promote explainability.[213] What matters is whether the system treats similarly situated individuals alike, and whether its outcomes reflect genuine need and natural variation, or whether they replicate bias through engineered thresholds and latent hierarchies.[214]

Individuals must be treated as ends in themselves and be afforded meaningful structural ways for contestation before harm occurs.[215] This goes beyond existing soft-law frameworks. As early as 2020, the Council of Europe adopted recommendations on the human rights impacts of algorithmic systems, affirming that individuals should be granted effective means to challenge determinations and decisions that affect them. But the proposal here departs from that view: it does not treat contestation as a matter of post hoc procedural redress. It insists, instead, on protections embedded into the structure of decision-making itself - ex ante, before harm is inflicted.²¹⁶ A parallel logic appears in *Goldberg v. Kelly*, where New York City terminated welfare benefits without giving recipients an opportunity to challenge the decision in advance.²¹⁷ The state argued it could cut off aid first and offer a hearing later.²¹⁸ The Court rejected this view.²¹⁹ It held that such a practice failed to satisfy the Due Process Clause.²²⁰ Due process, it ruled, requires timely notice and a hearing before the termination of essential benefits, particularly when those benefits concern basic subsistence.²²¹ "The fundamental requisite of due process of law," the Court declared, "is the opportunity to be heard at a meaningful time and in a meaningful manner."²²² To be meaningful, the hearing must precede deprivation and offer a genuine chance to contest the evidence on which the decision rests.²²³ By the same logic, the procedural protections proposed by conventional AI governance that activate only after an algorithmic system has acted are insufficient. What is needed is structural protection, built in from the outset, to secure a meaningful opportunity to be heard, at a meaningful time, and in a meaningful way.²²⁴

---

[209] 118 U.S. 356, 368-370 (1886).
[210] *Id* at 373.
[211] *Id.*
[212] *See* Chiraag Bains, *The Legal Doctrine that Will be Key to Preventing AI Discrimination*, BROOKINGS (September 13, 2024), https://www.brookings.edu/articles/the-legal-doctrine-that-will-be-key-to-preventing-ai-discrimination/.
[213] *See* Jessica Newman, *Explainability Won't Save AI*, BROOKINGS (May 19, 2021), https://www.brookings.edu/articles/explainability-wont-save-ai/ (arguing that explainability, while widely championed as a solution to AI's opacity, isn't sufficient to ensure trustworthy or just AI systems, because the current XAI methods are narrowly engineered for internal debugging and control, primarily serving the needs of developers – not users, communities, or regulators. And this has led to a mismatch between the ideal of explainability and its practice.)
[214] *See* Nicol Turner Lee, Paul Resnick, and Genie Barton, *Algorithmic Bias Detection and Mitigation: Best Practices and Policies to Reduce Consumer Harms*, BROOKINGS (May 22, 2019) (saying that algorithmic decision making falls short of our expectation because machines treat similarly situated people and objects differently); Stefan Feuerriegel, Mateusz Dolata & Gerhard Schwabe, *Fair AI*, 62 BUS. & INFO. SYS. ENG. 379-384 (2020) (individual fairness is based on the notion that similarly situated individuals should be treated in a similar way).
[215] *See* IMMANUEL KANT, FOUNDATIONS OF THE METAPHYSICS OF MORALS: AND WHAT IS ENLIGHTENMENT? 47 (Lewis White Beck Trans., 1959) (1785) ("Act so that you treat humanity, whether in your own person or in that of another, always as an end and never as a means only.")

The structural protections must extend to all persons. In *Plyler v. Doe*, the Supreme Court held that the Equal Protection Clause applies to "any person within the jurisdiction" of the United States, not only those with lawful status.²²⁵ Texas could not justify excluding undocumented children from access to public education based solely on immigration status.²²⁶ Constitutional safeguards follow personhood, not paperwork.²²⁷ That principle holds even in matters of war. In *Hamdi v. Rumsfeld*, the Court affirmed that a citizen designated as an enemy combatant retained the right to contest his detention before a neutral decision-maker.²²⁸ A state of war, the Court warned, is not a blank check for the President.²²⁹ Everyone subject to the coercive power of the state, regardless of legal status or alleged threat, must be afforded the right to be heard, to know the grounds of their classification, and to challenge it in a forum that offers genuine redress.

States must be held responsible for statutory and constitutional violations caused by the AI systems they deploy.²³⁰ As the Supreme Court mentioned, "the action of state courts and judicial officers in

---

²¹⁶ *See* Council of Eur., Recommendation CM/Rec (2020) 1 of the Committee of Ministers to Member States on the Human Rights Impacts of Algorithmic Systems 9, 13 (2020),
²¹⁷ 397 U.S. 254, 255 (1970)
²¹⁸ *Id* at 261 (the state and city officials contended that the combination of a post-termination "fair hearing" with the informal pre-termination review disposed of all due process claims).
²¹⁹ *Id.*
²²⁰ *Id.*
²²¹ *Id* at 267.
²²² *Id.*
²²³ *Id.* at 255 (holding that procedural due process requires a pretermination evidentiary hearing when public assistance payments to a welfare recipient are discontinued).
²²⁴ *See* Christopher L. Griff, Jr., Cas Laskowski and Samuel A. Thumma, *How to Harness AI for Justice*, 108 JUDICATURE INTERNATIONAL NO.1 (2024).
²²⁵ 457 U.S. 202, 215 (1982)
²²⁶ *Id* at 230.
²²⁷ *Id* at 210-211.
²²⁸ 542 U.S. 507, 509 (2004)
²²⁹ *Id* at 536.
²³⁰ *See generally* Kate Crawford and Jason Schultz, *AI Systems as State Actors,* 119 COLUM. L.REV. 1941 (2019) (arguing that AI vendors supplying systems for government decision-making should be considered state actors under

their official capacities is to be regarded as action of the State within the meaning of the Fourteenth Amendment," regardless of whether the underlying system is private.[231] In *Shelley v. Kraemer*, a Black family purchased a home in a St. Louis neighborhood governed by a racially restrictive covenant among white homeowners.[232] Although the covenant itself was private, the Court held that judicial enforcement of it violated the Equal Protection Clause.[233] By lending the force of law to a private exclusion, the state became the enforcer of racial discrimination.[234] The same logic applies to algorithmic systems. In spite of the private nature of their design, training, or ownership, once the state adopts, funds, or deploys them to carry out public functions, whether in housing, policing, education, or migration control, they operate as instruments of state authority. When such systems allocate benefits, deny access, or impose penalties, the constitutional threshold of state action is crossed; liability attaches to the state.

## C. Vertical and Horizontal Governance in a Consociational Order

Consociation offers an alternative to federalism by anchoring legitimacy in the representative authority of the governed.[235] Horizontally, it distributes authority across plural, self-constituting communities and sovereignty such as tribal nations, tenant unions, and municipalities; Vertically, it restrains federal overreach by affirming that while the federal government may encourage participation, it may not compel states' adoption or implementation of particular AI systems through administrative fiat. Both axes affirm an order in which authority is exercised with the consent of the governed.

### 1. Horizontally, tribal governments and local, community-based institutions should have the authority to authorize and reject AI systems within their existing domains.

Horizontally, tribal governments and community-based institutions such as municipalities must have the authority to authorize, condition, and reject the deployment of AI systems within their respective domains. The discussion here focuses primarily on tribal governments, whose jurisdictional boundaries are less widely understood,[236] whereas the authority of state and local governments to

---

the public function or joint participation tests of state-action doctrine; doing so is essential to closing the "accountability gap" and ensuring that constitutional and statutory protections apply to automated government systems)
[231] Shelly v. Kramer, 334 U.S. 1, 14 (1948)
[232] *Id* at 5.
[233] *Id* at 20.
[234] *Id.*
[235] *See generally* Daniel Elazar, *Federalism and Consociational Regimes*, 15 FEDERALISM AND CONSOCIATIONALISM: A SYMPOSIUM 17-34 (1985) (explaining that consociational democracy is defined by 1) coalitions of political leaders from all significant segments of a divided society, 2) segmental autonomy where each group governs its own affairs as much as possible, and 3) proportionality in representation and allocation, and 4) mutual veto rights to protect vital interests. These factors show an explicit effort to ground legitimacy in the representative authority of societal segments through a negotiated balance of segmental authorities. In addition, consociationalism involves processes of concurrent power sharing, institutionalized through the party system; such process is representative by nature, as legitimacy is mediated through elite actors who have effective authority over their communities. As Elazar observes, consociationalism tends to erode when representative authority declines.)
[236] *See* Stephen Cornell and Joseph P. Kalt, *American Indian Self-Determination: The Political Economy of a Policy that Works*, Faculty Research Working Paper Series at Harvard Kennedy School 1, 4 (2010) (saying that the status and genesis of American Indian Tribal Sovereignty are less widely understood)

regulate AI follows directly from constitutional federalism.²³⁷ Under the Tenth Amendment, powers not delegated to the federal government are reserved to the states, and by extension, to localities operating under state law.²³⁸ These powers include broad jurisdiction over health, safety, welfare, and public administration, which are all domains increasingly structured through algorithmic systems.²³⁹ Tribal governments, by contrast, are not creatures of the Constitution.²⁴⁰ They exist as distinct legal and political communities, recognized as domestic dependent nations that retain inherent sovereignty over their internal affairs.²⁴¹ This includes authority over membership, domestic relations, cultural preservation, and the governance of tribal property, ²⁴² for which the federal government has limited standing to intervene in these aspects.²⁴³ In *Santa Clara Pueblo v. Martinez*, a female tribal member challenged a rule that denied membership to the children of female members who married outside the tribe, while permitting it for similarly situated children of male members.²⁴⁴ The Court held that the Indian Civil Rights Act did not authorize federal courts to override such a rule on equal protection grounds.²⁴⁵ The decision reaffirmed that tribal self-government has the right to define community membership, even where those definitions depart from prevailing liberal norms.²⁴⁶ As the Court emphasized, tribal sovereignty is "a distinct attribute of Indian tribes," protected by federal law.²⁴⁷ They have power to make their own substantive law in internal matters.²⁴⁸ When tribal governments use algorithmic tools to distribute benefits, manage enrollment, administer justice, or preserve language and cultural memory, the deployment of those systems fall within the internal affairs of the tribe and must remain subject to tribal authority alone, even if their effects appear controversial to outsiders.[249]

---

²³⁷ *See e.g., Democracy: Co-Governance and the Future of AI Regulation*, 138 HARV. L.REV. 1609 (2025) (indirectly supporting the statement by noting that first, Americans trust state and local governments more than the federal government because those levels of government are likely more responsive to citizen concerns and it's easier to participate in them directly; second, it criticizes top-down regulation and repeatedly emphasizes that AI regulation should be brought closer to home, citing local knowledge, responsiveness, and adaptability; third, it makes clear that national rules and guidance can still reflect regional knowledge and preferences, and that consistent regulation can evolve through ways other than top-down dictates from Washington. This logic affirms that decentralized authority over AI -- including regulation by state and local governments - aligns with constitutional federalism, not in opposition to it; fourth, it distinguishes co-governance from localization, making clear that while the former often manifests locally, it draws legitimacy from democratic participation, not geographic jurisdiction per se. Local input and decentralized oversight is predicated on a federal constitutional structure that permits such jurisdiction)
²³⁸ U.S. CONST. AMEND. X
²³⁹ *See e.g., AI Adoption in the Public Sector: A New Study on Key Influencing Factors and Two Frameworks for Competencies and Governance,* EUROPEAN COMMISSION AI WATCH (November 25, 2024), https://ai-watch.ec.europa.eu/news/ai-adoption-public-sector-new-study-key-influencing-factors-and-two-frameworks-competencies-and-2024-11-25_en; *see also* Madeleine North, *6 Ways AI is Transforming Healthcare*, WORLD ECONOMIC FORUM (Mar 14, 2025), https://www.weforum.org/stories/2025/03/ai-transforming-global-health.
²⁴⁰ *See generally* Alisa Cook Lauer, *Dispelling Constitutional Creation Myth of Tribal Sovereignty, United States . Weaselhead*, 78 NEB. L. REV 162 (1999).
²⁴¹ *Id.*
²⁴² For a list of tribes recognized as political sovereigns by the federal government, *see* 25 C.F.R. § 83.5(a); *See also* COHEN'S HANDBOOK OF FEDERAL INDIAN LAW at ch.4-ch.7 (providing a detailed overview of these areas of American Indian Law, acknowledging tribal self-government and inherent sovereign powers, and all of the legal implications, including freedom from state laws).
²⁴³ *Id.*
²⁴⁴ 436 U.S. 49, 51 (1978)
²⁴⁵ *Id.* at 58.
²⁴⁶ *Id* at 55.
²⁴⁷ *Id.*
²⁴⁸ *See e.g.,* Roff v. Burney, 168 U.S. 218 (1897) (holding tribes determine their membership); Jones v. Meehan, 175 U.S. 1, 29 (1899) (holding tribes set up their own inheritance rules); United States v. Quiver, 241 U.S. 602 (1916) (holding tribes determine their domestic relations); Williams v. Lee, 358 U.S. 217 (1959) (tribes enforce laws in their own forums).

The authority of tribal governments over domestic relations and other internal affairs extends, in certain cases, to non-Indians.[250] In *United States v. Mazurie*, the Supreme Court upheld the criminal prosecution of two non-Indians, Martin and Esther Mazurie, who operated a bar—the Blue Bull—on fee land within the boundaries of the Wind River Reservation.[251] The tribal council, pursuant to an ordinance requiring both state and tribal licenses for liquor sales within Indian country, had denied the Mazuries a license.[252] When they nevertheless operated the bar, federal officers seized their inventory, and the Mazuries were charged under 18 U.S.C. § 1154 for introducing alcohol into Indian country.[253] The Court rejected the argument that Congress lacked authority to regulate the conduct of non-Indians on fee lands within reservations.[254] Citing *Williams v. Lee* and earlier precedents, it held that it was "immaterial that respondents were not Indians," because the conduct occurred within reservation boundaries and implicated core concerns of tribal governance.[255] Tribal authority, the Court emphasized, encompasses "the internal and social relations" of reservation life.[256] When Congress delegates regulatory power to tribes, it does so against the backdrop of tribes' inherent sovereignty, a power that persists even where tribal lands include non-Indian parcels. Extending this principle to digital infrastructure, it follows that when members and non-members alike engage with algorithmic systems administered by tribal governments, those systems fall squarely within the scope of internal tribal affairs and remain subject to tribal jurisdiction.[257]

Other institutions that do not possess inherent sovereignty, such as tenant councils, are nonetheless granted formal standing under federal law to participate in the governance of public housing. Under 24 C.F.R. § 245.100, tenants have an express right to organize,[258] and housing providers are required to "give reasonable consideration to concerns raised by legitimate tenant organizations"—a mandate that creates reciprocal obligations of consultation, access, and responsiveness.[259] Additionally, in public housing, resident councils are officially recognized as the "sole representative" of tenants, charged with advising on matters ranging from modernization and security to maintenance, resident selection, and recreational programming.[260] Drawing from that power, they compel collective bargaining, organize rent strikes, block evictions, and negotiate directly with landlords, lenders, and public agencies.[261]

---

[249] *See generally e.g.,* Adam Crepelle, *Tribes and AI: Possibilities for Tribal Sovereignty*, 25 DUKE L. & TECH. REV. 1 (2024) (arguing that the AI tools implemented by tribes for legal systems, healthcare, education and cultural efforts are extensions of tribal sovereignty; it emphasizes that such use must remain under tribal oversight); Ian Falefuafua Tapu & Terina Kamailelauli'i Fa'agau, *A New Age Indigenous Instrument: Artificial Intelligence & Its Potential for (De)Colonialized Data*, 57 HARV. C.R. – C.L.L REV 715 (2022).
[250] United States v. Mazurie, 419 U.S. 544, 558 (1975).
[251] *Id* at 546 – 547.
[252] *Id* at 548.
[253] *Id* at 550 – 551.
[254] *Id* at 552.
[255] *Id* at 558.
[256] *Id* at 557.
[257] *Supra* Note 265.
[258] 24 CFR § 245.100.
[259] 24 C.F.R. §245.105(b) (2024)

In this sense, when AI systems are introduced to manage tenant screening, allocate repairs, or flag nonpayment, the issue is not merely one of biased data or inadequate transparency.[262] It is that tenants, through their representative councils, should have a right to contest, shape, and in some cases veto the design and deployment of such systems.[263] Resident councils have the standing and functional expertise to evaluate whether algorithmic tools serve or undermine the collective interests of those who live under them. And housing authorities, as administrative agencies, have a duty to include tenant councils in these decisions, not as a courtesy, but as a matter of regulatory obligation and participatory justice.

Of course, other questions remain, such as the outer bounds of such nondelegation: How far can this authority extend before it breaches the constitutional limits on the delegation of public power? Can tenant councils or tribal governments lawfully veto or condition the deployment of algorithmic systems introduced by state agencies or federal programs? At what point does empowering non-state actors to authorize or block public deployments cross the line from democratic inclusion to unlawful delegation? What happens when a single municipality refuses to implement a system already adopted by state law enforcement? And what of other institutional actors, unions, school boards, churches? On what basis are some granted political standing while others are excluded, particularly in domains like migration or policing where their statutory authority is limited or unclear? These questions reach into deeper structural issues and fall beyond the scope of this paper. They will require their own treatment.

**2. Vertically, the federal government may offer incentives, condition funding, or invite state participation in algorithmic governance, but any such conditions must be grounded in clear statutory authority and stated with sufficient specificity to satisfy the Spending Clause. States retain the right to refuse participation, particularly where federal statutes intrude upon domains traditionally governed at the local level.**

---

[260] 24 C.F.R. § 964.11(a)
[261] *Supra* Note 257.
[262] *See generally* VIRGINIA EUBANKS, *Automating Inequality: How High-Tech Tools Profile, Police, and Punish the Poor* (St. Martin's Press 2018) (criticizing the imposition of algorithmic systems in public welfare and housing without community input).
[263] *Id.*

Vertically, the federal government may structure algorithmic governance by offering incentives, conditioning funding, or inviting state participation.[264] It may articulate broad policy goals, typically through statute, and delegate their implementation to state or local authorities.[265]

The conditions imposed by the federal government must be grounded in clear statutory authority and articulated with sufficient specificity to meet the requirements of the Spending Clause.[266] Vague aspirations will not suffice.[267] As the Court held in *Pennhurst State School & Hospital v. Halderman* (1981), Congress cannot bind states to new conditions not clearly stated in the legislation, because statutes enacted under the Spending Clause function "much in the nature of a contract": in exchange for federal funds, states must voluntarily and knowingly accept the terms imposed.[268] There can be no knowing acceptance, the Court emphasized, "if a State is unaware of the conditions or is unable to ascertain what is expected of it."[269] On that reasoning, statements like "people must be protected from unsafe or ineffective systems,"[270] or "algorithms should be equitable and non-discriminatory,"[271] as found in the AI Bill of Rights, are too abstract to create binding obligations on states.

Additionally, states can refuse to enact or implement federal statutes that deploy AI systems in domains traditionally governed at the local level.[272] The Tenth Amendment reserves to the states a distinct sphere of authority, particularly over matters such as public safety, education, housing, and local administration.[273] This is not a right belonging to governments alone; individuals, too, may assert federalism-based challenges when federal power intrudes upon these domains.[274] In Bond v. United

---

[264] *See* U.S. CONST. art. 1, § 8, cl.1; *see generally* South Dakota v. Dole, 483 U.S. 203 (1987) (Congress upholding the federal government's conditioning of highway funds on a state's adoption of a 21-year-old drinking age. The principle allows Congress to promote federal policy goals by attaching conditions to the receipt of federal funds, provided the conditions are in pursuit of the general welfare, unambiguous, related to the federal interest in the program, not independently unconstitutional, and not coercive). *See* U.S. CONST. art. VI, cl.1 (Congress may preempt state laws. States have an incentive to coordinate with federal guidance lest they find their own regulatory regimes displaced.); *See also* Printz v. United States, 521 U.S. 898 (1997); New York v. United States 505 U.S. 144 (1992) (the federal government may not compel state or their officials to enact or administer a federal regulatory program).
[265] *Id.*
[266] Pennhurst State School and Hospital v. Halderman, 451 U.S. 1, 16-17 (1981) (articulating Congress may fix the terms on which it shall disburse federal money to the States, and that the legitimacy of Congress' power to legislate rests on whether the State voluntarily and knowingly accepts the terms of the "contract".)
[267] *Id* (explaining that if a State is unaware of the conditions or is unable to ascertain what is expected of it, there can be no knowing acceptance. So, if Congress intends to impose a condition on the grant of federal moneys, it must do so unambiguously)
[268] *Id.*
[269] *Id. See also generally*, Charles C. Steward Mach. Co. v. Davis, 301 U.S. 548, 584 (1937).
[270] *See Blueprint for an AI Bill of Rights*, THE WHITE HOUSE, https://bidenwhitehouse.archives.gov/ostp/ai-bill-of-rights/ (last visited July 28, 2025).
[271] *Id* (setting forth the requirement for discrimination protections that one shall not face discrimination by algorithms and systems used and designed in an equitable way).
[272] as according to the anti-commandeering doctrine, the federal government may not compel states to legislate or execute federal law.
[273] U.S. CONST. AMEND. X
[274] *See* Bond v. United States, 564 U.S. 211, 222 (2011) ("the limitations that federalism entails are not therefore a matter of rights belonging only to the states… an individual has a direct interest in objecting to laws that upset the constitutional balance between the National Government and the States when they enforcement of those laws causes injury that is concrete, particular, and redressable.)

States (2011), the Court addressed precisely such an intrusion.[275] A woman was prosecuted under the federal Chemical Weapons Convention Implementation Act for poisoning her husband's mistress, conduct that would ordinarily fall under state criminal law.[276] The Court held that she had standing to challenge the statute under the Tenth Amendment, affirming that federalism protects individual liberty by denying any single government total jurisdiction over all matters of public life.[277] Individuals are also protected from such arbitrary overreach of congressional power beyond its constitutional limits.[278] Applied here, that principle would mean that if Congress enacts statutes such as the Federal Artificial Intelligence Risk Management Act of 2023 or the Artificial Intelligence Research, Innovation, and Accountability Act of 2024, and those statutes impose algorithmic systems in areas like housing, education, or local law enforcement, states may lawfully decline to carry them out. The federal government may not compel state agencies to administer its programs. Individuals, in turn, may have structural standing to challenge such overreach.

## D. The Right to Lawful Resistance

Compared to the two categories of covenantal and consociational order discussed above, the right to resist remains more conceptual at this stage. It refers to the individual's right and, at times, duty to resist tyranny, understood as the exercise of power beyond its legitimate bounds.[279] As a forward-looking principle, this right aims to coordinate the judgment and action of those who would restore legitimate authority.

The right to resist unjust tyranny is constitutionally embedded in the American tradition. It appears in both the Virginia Declaration of Rights and the Declaration of Independence.[280] In the latter, Jefferson notes that the colonists first exercised their right of petition—a form of peaceful address that stands as a predicate to rebellion.[281] Only after repeated injury and refusal does the sovereign forfeit his claim to rule, thereby triggering the right of resistance.[282] In Jefferson's words, he [King George] becomes "a Prince, whose character is thus marked by every act which may define a Tyrant, [and] is unfit to be the ruler of a free people."[283] Resistance becomes the rightful response to power that has severed its bond with the governed. The same logic informs the Second Amendment's opening clause: a well regulated militia is not merely a defense against foreign invasion, but a constitutional safeguard of a free polity.[284] Many state constitutions go further.[285] New Hampshire, North Carolina, and Tennessee declare that "the doctrine of nonresistance to arbitrary power and oppression is absurd, slavish, and destructive of the good and happiness of mankind."[286] The Maryland Declaration of Rights provides that the people may "reform the old or establish a new government" whenever liberty is manifestly endangered and no other means of redress remain.[287] Others dispense even with that condition. They affirm that the people may "alter, reform, or abolish" their government as a matter of first principle.[288] In each case, the structure holds: political power is not absolute. It is held on condition. When that condition is broken, resistance becomes a duty.

---

[275] *Id.*
[276] *Id.* at 214-215.
[277] *Id.* at 222.
[278] *Id.*
[279] *See* Ian Turner, *Resistance to Tyranny versus the Public Good: John Locke and Counter-Terror Law in the United Kingdom*, 20 DEMOCRACY & SEC. 321-346 (2024) (explaining a right to resistance to tyranny).
[280] THE DECLARATION OF INDEPENDENCE para. 2 (U.S. 1776) ("That whenever any Form of Government becomes destructive of these ends, it is the Right of the People to alter or to abolish it, and to institute new Government, laying its foundation on such principles and organizing its powers in such form, as to them shall seem most likely to effect their Safety and Happiness."); VIRGINIA DECLARATION OF RIGHTS art. III (1776), *reprinted in* 5 THE FOUNDERS' CONSTITUTION 3, 3 (Philip B. Kurland & Ralph Lerner ed., 1987) ("Whenever any government shall be found inadequate or contrary to these purposes, a majority of the community hath an indubitable, unalienable, and indefeasible right, to reform, alter, or abolish it, in such manner as shall be judged most conducive to the publick weal.")
[281] *Id.*
[282] *Id.*
[283] THE DECLARATION OF INDEPENDENCE para. 31 (U.S. 1776).

To say that this right is forward-looking is to highlight its self-enforcing character.[289] It enables those subject to power to recognize when a line has been crossed and to coordinate a response—whether through litigation, protest, or data strikes.[290] This dynamic is most clearly articulated in Barry Weingast's theory of the self-enforcing constitution, where he argues that a constitution binds only when citizens are willing and able to act in concert against violations.[291] When rulers transgress constitutional limits, their success depends on whether the public can recognize the breach and respond collectively.[292] If only one actor resists, they are likely to be punished or ignored. But when citizens converge in judgment, e.g., when they share an understanding that authority has overstepped -- resistance becomes possible.[293] By giving legal form to this right, the threshold for coordination is lowered.[294] The law signals not only that resistance is permitted, but when, where, and how it may rightly be carried out.

---

[284] U.S. CONST. AMEND. II. *See also* Steven J. Heyman, *Natural Rights and the Second Amendment*, 76 CHI. – KENT L. REV. 237, 247-49 (2000) (discussing Lockean influence on the Second Amendment).

[285] *See e.g.,* PA. CONST. art I, § 2 ("All power is inherent in the people, and all free governments are founded on their authority and instituted for their peace, safety and happiness. For the advancement of these ends they have at all times an inalienable and indefeasible right to alter, reform or abolish their government in such manner as they may think proper."); KY CONST. § 4 ("All power is inherent in the people, and all free governments are founded on their authority and instituted for their peace, safety, happiness and the protection of property. For the advancement of these ends, they have at all times an inalienable and indefeasible right to alter, reform or abolish their government in such manner as they may deem proper"); TEX. CONST. art. I, § 2 ("All political power is inherent in the people, and all free governments are founded on their authority, and instituted for their benefit. The faith of the people of Texas stands pledged to the preservation of a republican form of government, and, subject to this limitation only, they have at all times the alienable right to alter, reform or abolish their government in such manner as they may think expedient.")

[286] N.H. CONST. art. X ("Government being instituted for the common benefit, protection, and security, of the whole community, and not for the private interest or emolument of any one man, family, or class of men; therefore, whenever the ends of government are perverted, and public liberty manifestly endangered, and all other means of redress are ineffectual, the people may, and of right ought to reform the old, or establish a new government. The doctrine of nonresistance against arbitrary power, and oppression, is absurd, slavish, and destructive of the good and happiness of mankind."); Declaration of Rights and Other Amendments, North Carolina Ratifying Convention art. III (Aug. 1, 1788), *reprinted in* 5 THE FOUNDERS' CONSTITUTION, *supra* note 79, at 17 ("That government ought to be instituted for the common benefit, protection and security of the people; and that the doctrine of non-resistance against arbitrary power and oppression is absurd, slavish, and destructive to the good and happiness of mankind."); TENN. CONST. art. I, § 2 ("That government being instituted for the common benefit, the doctrine of non-resistance against arbitrary power and oppression is absurd, slavish, and destructive of the good and happiness of mankind.").

[287] MD. CONST. of 1776, Declaration of Rights art. IV, *reprinted* in 4 SOURCES AND DOCUMENTS OF UNITED STATES CONSTITUTIONS 372, 372 (William F. Swindler ed., 1975). *See generally* CHRISTIAN G. FRITZ, AMERICAN SOVEREIGNS: THE PEOPLE AND AMERICA'S CONSTITUTIONAL TRADITION BEFORE THE CIVIL WAR 13 (2008), at 24 (discussing New Hampshire borrowing Maryland's language).

[288] *See* FRITZ, *id* at 24 (discussing Pennsylvania).

[289] *See generally* Sonia Mittal and Barry R. Weingast, *Self-Enforcing Constitutions: With an Application to Democratic Stability*, 29 J.L. ECON. & ORG. 278-302 (2013) (explains that what it means to be self-enforcing is the constitutional order doesn't wait passively to be violated and then vindicated. It anticipates breach, structures incentives, and pre-commits actors to constraint. As the article puts it, "the Consensus condition addresses the fundamental coordination problem underlying democracy, namely, whether citizens have the ability to act in concert against political leaders who transgress constitutional rules… by defining what actions constitute transgressions, constitutions help citizens coordinate against potential violations.")

Such coordination differs from the right to contestation, though both aim to challenge algorithmic decisions. As Kaminski and Urban explain, the right to contestation is a procedural entitlement grounded in due process.[295] It allows individuals to challenge significant decisions made by or with the assistance of automated systems—but only after those decisions have been rendered and the harm has already begun to take effect.[296] It is retrospective, remedial, and individualized. Coordination, by contrast, is structural. It depends on a public capacity to recognize when systems overstep their bounds and to organize a collective response before the harm becomes entrenched. Contestation operates within the existing legal order; coordination signals that that order may no longer bind.[297]

---

[290] *Id* (emphasizing that a constitution must clearly define what counts as a transgression, because recognition is a prerequisite for coordinated action).
[291] *Id.*
[292] *Id.*
[293] *Id.*
[294] This is because, sometimes the ruler may desire to impose limitations on themselves, as such behavior, under some circumstances, promote economic development and might even secure political support. *See e.g.,* Daron Acemoglu & Simon Johnson, *Unbundling Institutions*, 113 J.POL. ECON. 949, 953 (2005) (suggesting that institutions with "greater constraints on politicians" are crucial for economic growth); Rafael La Porta et al., *Judicial Checks and Balances*, 112 J. POL. ECON. 445 (2004) (finding an empirical link between judicial constraints and economic freedom and property rights); Paul G. Mahoney, *The Common Law and Economic Growth: Hayek Might Be Right*, 30 J. LEGAL STUD. 503 (2001) (finding that the common law tradition of a free market and limited government is associated with economic growth); Douglass C. North & Barry R. Weingast, *Constitutions and Commitment: The Evolution of Institutions Governing Public Choice in Seventeenth-Century England*, 49 J. ECON. HIST. 803 (1989) (demonstrating how legal limitations on rulers' arbitrary power in early capitalist Europe increased the legal security and predictability of external lenders who were protected by law from the seizure of their capital, and by reducing the risks associated with lending, capital became more readily accessible, economic growth increased, and the relative positions of countries where sovereigns had limited themselves improved markedly).
[295] *See generally* Margot E. Kaminski and Jennifer M. Urban, *The Right to Contest AI*, 121 COLUM. L. REV. 1957 (2021).
[296] *Id.*
[297] For example, while laws such as GDPR Article 22(3) entitle individuals to obtain human intervention, to express their point of view, and to contest the decision when subject to automated processing, coordination rights would instead protect the conditions under which groups may form judgments, share information, and act together in response to structural encroachments. They but mark a shift from the procedural logic of appeal to the political logic of collective self-government.

# PART III. CONCLUSION

This Article has argued that the dominant paradigms of AI governance—whether the European model of rights-based risk management or the American model of decentralized federalism—fail to confront the foundational question of legitimate authority. Both offer techniques of containment rather than frameworks of rule. The EU model relies on procedural rights to stabilize technological disruption, exporting its legal architecture under the guise of moral universality. The U.S. model valorizes subnational experimentation but leaves the source of authority ambiguous, as if innovation alone could substitute for institutional clarity. In both, the result is the same: governance without grounding, administration without consent.

What is needed is not a better rights regime or more granular regulation, but a constitutional reconstruction at a higher level of law. This Article has proposed one, rooted in three principles: covenant, consociation, and resistance. Covenant reasserts that all rightful authority must be conferred by those subject to it. Governance begins not with technical capacity but with the founding act of mutual obligation and participatory consent. Consociation affirms that political life is inherently plural. AI systems operate across jurisdictions, communities, and domains of meaning, and must be governed accordingly, through representative institutions that speak for those they affect. Resistance reintroduces the structural right to withhold obedience from systems that overstep their bounds. Not as rebellion, but as a lawful act of fidelity to the original terms of public power.

This framework is not nostalgic or historical. It is constitutional. It does not seek to return to some pre-technological past, but to restore the conditions under which technological governance may be exercised lawfully. As AI systems increasingly make decisions once reserved to human judgment—allocating benefits, determining eligibility, shaping legal outcomes—the question is no longer whether they are effective or efficient. The question is whether they rule, and if so, by what right. Until that question is squarely faced, AI regulation will remain structurally incomplete. The work of building legitimacy, of reconstructing the terms on which power is exercised and contested, must begin again, from below.